\documentstyle[12pt,aasms4,epsf]{article}
\newcommand{\ho}{H$_2$O}
\newcommand{\ccm}{cm$^{-3}$}

\newcommand{\kms}{km\thinspace s$^{-1}$}
\newcommand{\et}{{\it et\thinspace al.}}

\newcommand{\simless}{\mathbin{\lower 3pt\hbox
     {$\rlap{\raise 5pt\hbox{$\char'074$}}\mathchar"7218$}}} 
\newcommand{\simgreat}{\mathbin{\lower 3pt\hbox
     {$\rlap{\raise 5pt\hbox{$\char'076$}}\mathchar"7218$}}} 

\lefthead{Strelnitski \et.}
\righthead{\ho\ Masers and Supersonic Turbulence}

\begin{document}
\baselineskip 14pt

\title{\ho\ MASERS AND SUPERSONIC TURBULENCE}

\author {V. Strelnitski}
\affil {Maria Mitchell Observatory,  3 Vestal St., Nantucket, MA 02554,\break
and New Mexico Institute of Mining and Technology}
\author { J. Alexander}
\affil{New Mexico Institute of Mining and Technology, Physics Department,
Socorro, NM 87801}
\author {S. Gezari}
\affil {Brown University, Box 4186, Providence, RI  02912, \break
and Maria Mitchell Observatory}
\author{B.P. Holder}
\affil {Wesleyan University, Middletown, CT 06459,\break
 and Maria Mitchell Observatory}
\author {J.M. Moran and M.J. Reid}
\affil {Harvard-Smithsonian Center for Astrophysics, 60 Garden Street,
Cambridge, MA 02138}
\begin{abstract}
We use unpublished and published VLBI results to investigate the
geometry and the statistical properties of the velocity field traced by
\ho\ masers in five galactic regions of star formation --- Sgr B2(M),
W49N, W51(MAIN), W51N, and W3(OH).  In all sources the angular
distribution of the \ho\ hot spots demonstrates approximate
self-similarity (fractality) over almost four orders of magnitude in
scale, with the calculated fractal dimension $d$ between $\approx 0.2$
and 1.0. In all sources, the lower order structure functions for the
line-of-sight component of the velocity field are satisfactorily
approximated by power laws, with the exponents near their classic
Kolmogorov values for the high-Reynolds-number incompressible
turbulence. These two facts, as well as the observed significant excess
of large deviations of the two-point velocity increments from their
mean values, strongly suggest that the \ho\ masers in regions of star
formation trace turbulence. We propose a new conceptual model of these
masers in which maser hot spots originate at the sites of ultimate
dissipation of highly supersonic turbulence produced in the ambient gas
by the intensive gas outflow from a newly-born star. Due to the high
brightness and small angular sizes of masing hot spots and the
possibility of measuring their positions and velocities with high
precision, they become a unique probe of supersonic turbulence. 
\vskip24pt\noindent
{\it Subject headings:} masers --- turbulence --- ISM: jets and outflows
\end{abstract}
%
%
\section{Introduction}

The idea of an energy cascade through a hierarchy of scales (Richardson
1922; Kolmogorov 1941a; Obukhov 1941), the phenomenological theory of
turbulent energy dissipation (Kolmogorov 1941a,b, 1942), and the
experimental and theoretical results related to the intermittency of
turbulent velocity fields (see Frisch 1995 for references) are the
cornerstones of the present understanding of incompressible
turbulence.  Much less is known about compressible (supersonic)
turbulence.  Analytical approaches and laboratory experiments are still
limited to low Mach numbers, M$\,\approx 1$, and the results of
numerical simulations may depend crucially on the adopted boundary and
initial conditions.

Von Weizs\"acker (1951) was the first to point out that the
interstellar medium (ISM) is characterized by very high Reynolds
numbers and high Mach numbers, and thus highly supersonic turbulence
should be its typical state. Much work using the ISM to study
supersonic turbulence has since been attempted (see, for example,
Dickman 1985; Scalo 1987; Falgarone \& Phillips 1990; Elmegreen 1993;
Franco \& Carraminana 1999).  However, at the relatively large scales
considered so far, $10^{21} \to 10^{17}\,$cm, the ISM may not be an
adequate laboratory for studying supersonic turbulence in the
traditional sense of the term ``turbulence.'' The initially
hypothesized turbulent energy cascade from the largest to the smallest
scales of the ISM, initiated by the differential rotation of the Galaxy
(von Weizs\"acker 1951; Fleck 1981) is an obvious idealization. Even if
the differential rotation works as the major energy source for the
observed turbulence in the neutral gas at the perifery of galaxies
(Sellwood, \& Balbus 1999), it should be disrupted by the powerfull
injection of energy from supernovae and stellar winds at intermediate
scales in the major galactic disk (Spitzer 1978). ISM turbulence  at
large and intermediate scales is also complicated by the effects of
self-gravitation.

It has recently been recognized that the 1.35-cm wavelength \ho\ masers
[see reviews on masers in Elitzur (1992); Clegg and Nedoluha (1993);
Migenes and Reid (2002)] may be promising tools for the study
of ``Kolmogorov-type'' supersonic turbulence.  VLBI studies of the
proper motions of several bright \ho\ maser sources associated with
newly-born stars have revealed expansion of the clusters of maser spots
--- participation in gas outflows from these stars (see Anderson and
Genzel 1993 for a review).  This type of regular motion had been
theoretically predicted before its discovery (Strelnitski \& Sunyaev
1973).  In some cases, including W49N, there are also indications of
another regular component in the velocity field revealed by the masers
--- rotation (Reid \et\ 1988, Gwinn \et\ 1992).  However, besides these
regular components, VLBI measurements indicated the presence of a
residual {\it random} component of motion.  Typically, approximating
the proper motion vectors by a simple model of expanding and rotating
gas leaves a residual dispersion of $\approx 15\,$\kms\ per axis, which
is considerably larger than the errors of these observations (Reid
\et\ 1988). This value corresponds to $\approx \sqrt3\times 15\approx
26\,$\kms\ for the total velocity vector and can be considered the
characteristic turbulent velocity dispersion at the largest spatial
scale covered by the maser cluster.

Walker (1984), using the VLBI maps of the \ho\ source in W49N obtained
by Walker, Matsakis \& Garcia-Barreto (1982), demonstrated that both
the two-point velocity increments and the two-point spatial correlation
function show power-law dependencies on maser spot separation.  This
behavior is typical of a turbulent flow, although Walker did not favor
the turbulence interpretation.  Gwinn (1994) carried out a similar
statistical analysis using better VLBI results for W49N obtained by
Gwinn \et\ (1992). He confirmed the power-law dependency of velocity
dispersion and spatial density of masing spots on spatial scale and
ascribed this behavior to turbulence.

This paper summarizes a series of our studies of \ho\ masers as tracers
of supersonic turbulence in regions of star formation.  Preliminary
results were reported by Strelnintski \et\ (1998), Holder \&
Strelnitski (1997), Gezari (1997) and Gezari, Reid, \& Strelnitski
(1998). Section~2 presents our VLBI results for the \ho\ maser source
in Sgr B2(M). The geometrical properties of supersonic turbulence
revealed by \ho\ masers in this and four other sources are presented in
Section~3,  and the statistical properties of the velocity field traced
by the masers in Section~4.  In Section~5, proceeding from the
hypothesis that the \ho\ masers adequately probe the velocity field of
turbulence, we discuss implications of our statistical results for the
theory of supersonic turbulence. In Section~6 we describe a new
conceptual model of \ho\ masers in regions of star formation based on
the surmised connection of maser pumping with the sites of ultimate
dissipation of turbulent energy.  Conclusions are summarized in
Section~7.

%
%
\section{VLBI Observations of Sgr B2(M)}
The observations of Sgr B2(M) were conducted in 1986 on 23 January, 26
February, 27 March, and 26 April, as part of a campaign to measure
proper motions of these H$_2$O masers.  Four telescopes spanning the
U.S. were used: the Haystack 37-m telescope in Westford, MA, the
NRAO\footnote{NRAO is a facility of the National Science Foundation,
operated under cooperative agreement by the Associated Universities,
Inc.} 43-m telescope in Green Bank, WV, one 25-m telescope of the VLA
near Socorro, NM, and the OVRO 40-m telescope in Big Pine, CA.  The Mk
III recording system was used with four 2-MHz bands covering the LSR
velocity ranges of -40 to -14, +18 to +43, +43 to +69, and +68 to +94
km~s$^{-1}$, assuming a rest frequency of 22235.08 MHz for the H$_2$O
$6_{16} -- 5_{23}$ transition.  The recorded data were
correlated at the Mk III processor at Haystack Observatory in a mode
which yielded 56 (uniformly weighted) spectral channels, each 35.71 kHz
or 0.48 km~s$^{-1}$ wide.

The data were edited, calibrated, and imaged following the same general
procedures as described in Reid \et\ (1988) for the source Sgr B2(N).
The synthesized interferometer beam was approximately 2.4 by 0.4 mas
FWHM, elongated in the north-south direction, owing to the low
declination of Sgr B2.  Compact maser spots with flux densities ranging
between 135 and 0.4 Jy were detected across a field of approximately 2
by 2 arcseconds.  The positions of the maser spots were obtained by
fitting a circular Gaussian brightness model for each spectral channel
independently using the AIPS task IMFIT.
%
%
\section{The Geometry of \ho\ Masers}
Figure~1 shows a series of decreasing spatial scales for our VLBI map
of Sgr B2(M).  The scale changes by almost four orders of magnitude ---
from $\simgreat 1\,$arcsec down to $\simless 1\,$mas --- which, at the
assumed distance of 8~kpc (Reid 1993), corresponds to a range of linear
scales between, roughly, 10,000 and 1~A.U. As is typical of fractal
dust-like structures, the distribution of masers looks qualitatively
the same on all scales, with evident clustering.

On all maps of Figure~1, except the last one, dot sizes are larger than
the typical observed size of an individual maser spot. Dots on the last
map show measured positions of spectral channels. Since a spectral
channel ($0.48 \,$\kms) is narrower than a typical spectral width of a
single spot ($\simgreat 0.8\,$\kms), a dot on the last map typically
represents the position and velocity of only a part of an individual
spot. We call the smallest groupings of heavily blended individual
spots ``minimal clusters.''  The dots on the map only approximate the
extension of and the velocity dispersion in these minimal clusters.
Inspection of the available data indicates that a typical size of a
minimal cluster is a few A.U. The velocity dispersion within minimal
clusters varies from $\approx 1$ to 5 \kms.

We used two methods to estimate the fractal dimension of the spatial
distribution of \ho\ masers in SgrB2(M): the ``density-radius'' and the
``box-counting'' methods (see e.g. Crownover 1995; Feder 1988).  The
density-radius measure is based on the generalization of the mass {\it
versus} radius relation for objects of integer dimension,
%
\begin{equation}
     M \propto r^d\;, 
\end{equation}
where $d$ is the dimension of the object.  Equation~(1) can be used as
a general definition of the dimension of an object, including objects
whose average density changes in a self-similar way with changing scale
(e.g.  Mandelbrot 1982). For these objects (fractals) $d$ is
non-integer. Average density, $\rho$, within a given volume $V$ is
$M/V$. Therefore,
%
\begin{equation}
    \rho = M/V \propto{{ r^d}\over {r^{d_0}}} = r^{d-d_0}\;,
\end{equation}
where $d_0$ is the dimension of the supporting space, e.g. $d_0 = 2$ for
a plane.  If the dimension of the object equals the dimension of the
support, $\rho$ equals a constant. If not, $\rho$ is a function of $r$.
The steepness of this function depends on $d$.  

A more practical procedure results from differentiating Equation~(1), 
which gives
%
\begin{equation}
\sigma \equiv {{{\rm d}M}\over {{\rm d}V}} \propto r^{d-d_0}\;,
\end{equation}
or
%
\begin{equation}
d = {{{\rm d(log}\,\sigma)}\over {{\rm d(log}\,r)}} + d_0\;.
\end{equation}
Determining the dimension is thus reduced to measuring the slope of
${\rm log}\,\sigma$ {\it versus} log$\,r$. In the case of a point set
like ours, ``density'' means ``number density.''

In our numerical procedure, $\sigma$ is calculated (for a discrete set
of $r$ values) as the surface density of companions to a given maser
hot spot at angular separation $r$, averaged over all maser spots:
%
\begin{equation}
\sigma(r) = {{<{\rm number\_of\_points}\,(r, r+\delta r)>}\over 
{\pi[2r\delta r+(\delta r)^2]}}\;. 
\end{equation}
The same procedure was used by Walker (1984) and Gwinn (1994) for
demonstrating self-similar clustering of \ho\ masers in W49N. However,
these authors did not relate $\sigma (r)$ to the fractal dimension of
the source. Larson (1995) uses a similar procedure to obtain a fractal
dimension for a young stellar association in Taurus.

We tested our numerical procedure by obtaining the density-radius
fractal measure of a simple straight line and of the classical
mathematical fractal, a Sierpinski triangle.  The measured dimensions
of 1.00 and 1.58 were in excellent agreement with their theoretical
values $d = 1.000$ and $\approx 1.585$, respectively. 

The box-counting measure associates a fractal object's dimension $d$
with the number $N$ of boxes of side length $l$ needed to cover the
object (Crownover 1995):
%
\begin{equation}
N(l) \propto l^{-d}\,. 
\end{equation}
The graph of log$\,N(l)$ {\it versus} log$\,l$ is a straight line,
having slope $-d$. We used the following computational algorithm. The
square plane of minimal side length $L$ containing the whole object is
divided into $2^2$ equal squares of side length $L/2$.  The number of
these squares containing one or more points making up the object is
determined and stored.  Each non-empty square of side length $L/2$ is
subdivided again into 4 squares of equal area; the number of non-empty
squares of side length $L/4$ is determined and stored, etc.  The
procedure is repeated down to some minimal side length of sub-squares;
minimal side length is determined by the characteristic length of the
smallest features of the object.  Logarithm of the number of non-empty
squares {\it versus} logarithm of their side length is plotted, and the
slope of the straight line fitting the data points is measured; this
slope is equal to $-d$. This numerical procedure was also tested with a
straight line and a Sierpinski triangle. The measured values of
dimension were again in agreement with the theoretical values.

We used equation~(4) to determine the fractal dimension of the 2D
projected ($d_0 = 2$) \ho\ masers in the four observations of SgrB2(M).
The values of the fractal dimension for the four observations and for
their average are indicated in the corresponding panels of Figure~2.
The average value is $d_2 \approx 0.44\pm 0.07$. The box-counting
result for the average of the four observations is shown in Figure~3.
The ensuing fractal dimension, $d_2 \approx 0.21\pm 0.02$, is
noticeably lower than that obtained from the density-radius plot.

For both measures, a single linear fit is a satisfactory first
approximation; the standard deviation of the residuals to the fit does
not surpass $\pm 10\%$ for each epoch of observations and for the
combined fit.  However, notable deviations from a single power-law
approximation can be seen in Figures~2 and 3.  For example, a higher
fractal dimension for the largest scales is evidenced by the steeping
slope of the points in Figure~3. Some deviation from the linear
dependence is seen in Figure~2, between log (separation) $\approx -2$
and $-3$. It may indicate some depression of clustering at the scales
around 0.003 arcsec ($\sim 10^{14}\,$cm for this source).

The origin of the systematic difference between the two applied fractal
measures is unclear.  It may be rooted in technical particularities of
the methods.  Some practical problems in the application of the
box-counting method are discussed in Gouyet (1996, Section 1.4.4). This
discussion indicates that derived fractal dimensions should be correct
within a factor of two.  Given this uncertainty and the range of the
derived values of slopes in Figures~2 and 3, we estimate the fractal
dimension of the observed cluster as $d_2 \approx 0.3 \pm 0.2$.

Gwinn (1994) performed a statistical analysis of the VLBI maps of
W49/\ho\ obtained by Gwinn \et\ (1992). He demonstrated a power-law
dependence of the number density of neighbors on their separation.
However, he used the one-dimensional projection of the distance and did
not interpret his results in terms of fractals. To make results for
W49N comparable to those for SgrB2(M), we applied the density-radius
fractal measure to the two-dimensional spatial distribution of the
\ho\ maser spots in W49N, using the VLBI positions published by Gwinn
\et\ (1992). We have applied this measure to three more sources --- two
with the published VLBI results: W51(MAIN) and W51N (Genzel et al.
1981; Schneps et al. 1981) and one source, W3(OH), for which we used
our unpublished VLBI coordinates of the maser spots (the corresponding
map of the source was published --- Alcolea et al. 1992).

The results for all sources are shown in Figure~4. A power-law
dependence is a good approximation for all of them and it gives a low
fractal dimension, $\simless 1$, for all the observed sets of maser
spots as projected on the sky.

Given the small angular dimensions of maser clusters ($\sim
1\,$arcsec), their projection on the sky is essentially an orthogonal
projection.  Therefore, the dimension $d_3$ of the real fractal,
residing in 3-space, coincides with the dimension, $d_2$, of its 2D
projection if $d_3 = d_2<2$ (e.g. Falconer 1990). We have demonstrated
above that this condition is fulfilled for \ho\ masers.  Thus, we
conclude that the fractal dimension of \ho\ clusters in all five of the
sources we have studied is low, $d_3 \simless 1$.

This conclusion is new, although  Walker (1984) and Gwinn (1994)
obtained results that can be converted to estimates of fractal
dimension.  Walker (1984) obtained a high negative value of the power
index ($\approx - 1.1$) in the power-law approximation of the
density-radius dependence for W49N/\ho.  This corresponds to a fractal
dimension $d_3 = d_2 \approx 0.9$. Gwinn (1994) reduced his two-point
correlation analysis of \ho\ masers in W49N to the 1D projection of the
observed map on the $x$ axis.  He obtained a power index, $\gamma_1
\approx - (0.2\to 0.3)$, for a large interval of scales. Although he
did not connect this result with a fractal dimension, we note that for
a 1D-projected fractal, $d_1 = \gamma_1 + 1$, where $d_1$ is the
fractal dimension of the 1D projection of the real fractal residing in
3-space.  Thus, Gwinn's result corresponds to $d_1 \approx 0.7 - 0.8$.
Since $d_1 < 1$, the same fractal dimension is ascribed to both the
2D-projection and the real fractal residing in 3-space. Thus, both the
Walker's (1984) and the Gwinn's (1994) results support our conclusion
that the fractal dimension of \ho\ clusters is $\simless 1$.

To better appreciate the fractal distribution of \ho\ masers, it is
instructive to compare it with models of homogeneously distributed
random points.  In our numerical model we created 90 points randomly
and uniformly distributed in a thin spherical shell and then projected
this distribution onto a plane. A drastic difference between the model
distribution and the observed maser distribution can be seen visually
(a lack of clustering in the model distribution) and is confirmed by
the measured spatial dimension of the model point sets. As anticipated,
both box-counting and density-radius methods gave $d\approx 2$ for the
random, homogeneous cluster of model dots, to be compared with
$d\simless 1$ for the observed clusters of maser spots.

Other types of masers should also be tested for possible fractal
structure. At least some of them do not seem to have such structure.
For example, the OH masers associated with regions of star formation do
not show self-similar spatial distribution, rather they demonstrate
strong clustering on one scale, $\sim 10^{15}\,$cm (Reid \et\ 1980).
These masers form just outside an expanding ultra compact HII region and
would not be expected to have a turbulent structure of the same kind as
the \ho\ masers that are due to the shear between a stellar wind and
surrounding gas (see Section~6).
%
%
\section{Statistics of The Velocity Field}
We investigated two statistical properties of the velocity field traced
by \ho\ masers in the same five sources: (1) the low-order two-point
velocity structure functions, and (2) the probability distribution for
the deviations of the two-point velocity increment from its mean value
at different spatial scales.
%
\subsection{Two-Point Velocity Structure Functions}
Most statistical studies of the kinematics and structure of
interstellar gas using masers as probes have so far been limited to one
velocity component (the line-of-sight) and two coordinates on the
celestial sphere. Owing to the smallness of maser sources (a whole
source is only $\sim 1\,$arcsec across), the two spherical celestial
coordinates are, with high precision, approximated by rectangular
Cartesian coordinates.  We assume, as all previous authors implicitly
did, that if a power-law scaling relation takes place for velocity
vectors in 3D space, the same relation, with the same exponent, holds
for the dependence of the line-of-sight component of velocity on
projected distances. This is a reasonable assumption if the velocity
field is isotropic. It is analogous to the well-known use of the
longitudinal velocity component in incompressible turbulence studies
(see e.g. Frisch 1995).

Statistical analysis of the velocity field probed by \ho\ masers has
previously been performed for W49N on two independent sets of data
(Walker 1984, Gwinn 1994). We discuss here only the low-velocity
\ho\ maser spots ($\approx \pm 20\,$\kms\ from the systemic velocity),
which are more likely connected with the ``Kolmogorov-type'' supersonic
turbulence than the high-velocity spots (section~6).  Walker (1984) did
not see an explicit dependence of two-point velocity increments on
point separation for low-velocity features, shown in his Figure~8. One
can interpret this graph as a power law with the exponenent $q \simless
0.2$. Gwinn (1994) found $q \approx 0.33\pm 0.01$ for the dependence of
the median velocity differences on the 1D projection of the maser pair
separation.  One can conclude from these two studies that the value of
$q$ for the two-point correlation function in W49N/\ho\ does not
surpass 1/3.

For each of the five \ho\ sources (Section~3) we calculated the structure 
functions   
%
%
\begin{equation}
D_{\alpha}(l) \equiv \langle |v({\bf r}) - v({\bf r} + {\bf 
l})|^{\alpha}\rangle\;.
\end{equation}
for low values of the order of the function, $\alpha = 1\to 3$.  In
equation~(7) the vectors {\bf r} and {\bf l} determine the positions of
the two points in the plane of the sky, $v$ designates the
line-of-sight velocity and $l$ the 2D projection of the linear distance
between the points.  Our calculation procedure is as follows. The whole
range of maser spot angular separations (up to four orders of
magnitude, in both sources) is divided into $N$ bins, with
logarithmically increasing bin size and thus logarithmically increasing
separation between bin-centers. Using the VLBI relative position data,
the procedure selects all pairs within a given separation bin and
calculates one of the functions (eq.[7]). This procedure is repeated for
every separation bin and the resultant averages of the powers of
velocity differences are plotted as a function of separation in a
log-log graph.  A least-squares fit of a straight line to the points on
the graph is then performed to obtain a power-law exponent, $q$, and
its uncertainty (one standard deviation).

Figure~5 gives the results for Sgr~B2(M), for $\alpha = 1$. The figure
shows the results for the four epochs of observations (Section~2), as
well as their average. The data are satisfactorily approximated by a
power law.  We obtain $q = 0.31\pm 0.03$, when the whole set of data is
used.  When data for the largest scales (log $|\Delta\theta\,{\rm
(arcsec)}| \ge -0.5$) are excluded (to avoid possible edge effects),
the value of $q$ changes insignificantly:  $q = 0.34\pm 0.03$. The
second and third order structure functions have the exponents of
$0.62\pm 0.06$ and $0.93\pm 0.09$, respectively. We note that for all
three structure functions the power-law exponents are close to
their classic Kolmogorov values, which are 1/3, 2/3 and 1.0 for $\alpha =
1,\;$2 and 3, respectively.

The results for $\alpha = 1$ for other sources are presented in
Figure~6.  The power-law approximation gives the values of $q$ close to
1/3 for all the sources except for W3(OH), where it is significantly
lower ($0.19 \pm 0.03$). In this source, however, the VLBI map reveals
a strong regular component of motion (strongly collimated bipolar
outflow), which should significantly influence the results when the
whole VLBI map is considered for the statistical analysis. To decrease
the influence of the regular velocity component, we obtained the first
order structure function for only one of the two lobes of the bipolar
outflow. In this case, the regular component of the relative velocities
 should be minimal, and one can anticipate that the bulk of the
relative motions of the condensations will be due to turbulence.  The
result is shown in Figure~7; the value of $q$ (0.30) is now much closer
to the Kolmogorov value.

With all sources displaying the low-order structure functions
close to Kolmogorov's, one might wonder how likely it is that this is simply
coincidental. Can regular, non turbulent velocity fields, such as
expansion and/or rotation, produce the observed power-law dependence of
the velocity increments on spatial scales, with the power index close
to 1/3?

In order to answer this question we applied the same statistical
analysis to the results of numerical modeling that simulated regular
motions of the maser spots only. 90 model dots were randomly and
uniformly distributed in a thin spherical shell. Three types of regular
motion were considered: (1) radial expansion; (2) rotation around an
axis perpendicular to the line of sight; and (3) expansion plus
rotation.  In the last case we varied the ratio of the absolute values
of the expansion velocity and the velocity of rotation on the equator.

The results of the (line-of-sight velocity) $\it versus$ (dot
separation) correlation analysis are shown in Figure~8. The long
straight line on the plots shows the slope 1/3 for reference. The
quickly growing dispersion of the data points at smaller scales, seen
on all the plots, is due to the uniform, non-fractal distribution of
the points --- a lack of clustering, and thus a poor statistics, at
smaller scales. This large dispersion makes a linear fit beyond about
1.5 orders of magnitude from the largest scale meaningless.  The slope
of the fitting line, drawn in this limited interval, changes from about
zero for the case of pure expansion ($\approx 0.08\pm 0.02$, for the
specific realization shown in Figure~8) to approximately unity for
the case of pure rotation ($1.01\pm 0.01$ in the example shown). All
intermediate values of the slope can be achieved by combining expansion
and rotation in a due proportion (the four lower panels). In
particular, the ``Kolmogorov'' value 1/3 is achieved when the ratio of
the velocity of rotation on the equator to the velocity of expansion is
$\approx 3$.

It is quite improbable, however, that this {\it ad hoc} combination of
kinematic parameters, plus the same orientation of the axis of rotation
is realized in all the sources under study. In all the published models
of the observed proper motions and radial velocities of \ho\ masers the
deduced model ratio of the expansion-to-rotation velocities is less
than one. In W49N these velocities are almost equal --- 17 and 16 \kms,
respectively (we consider here only the low-velocity componenent of
expansion; see Section~6). It is seen from Figure~8 that such a low
ratio of expansion-to-rotation should produce a flatter slope than 1/3
at large scales.  It is noteworthy that flattening of the slope of the
two-point correlation function does indeed appear at larger scales for
all the sources in Figure~6.  This reveals the contamination of the
statistical properties of the turbulent component of motion by the
regular component. The role of the regular component(s) of the velocity
field relative to the turbulent component drops with the decreasing
spatial scale, and it is remarkable how effectively the smaller scales
``compensate'' for the flattening at the larger scales in Figure~6 and
force the average slope to tend to its Kolmogorov value.

%
%
\subsection{Statistics of Deviations from the Mean Velocity Increment}
An inherent manifestation of terrestrial turbulent flows is {\it
intermittency} --- the spatial and temporal inhomogeneity of the
velocity field.  Intermittency results in enhanced, higher than
Gaussian, probability of large deviations of the two-point velocity
increments from their average value at a given spatial scale.
Deviations from a Gaussian distribution have been observed in
laboratory and atmospheric incompressible turbulent flows [Dutton and
Deaven (1969), van Atta and Park (1972)]. Falgarone and Phillips (1990)
and Falgarone, Phillips, and Walker (1991) attributed the broad
(broader than Gaussian) wings of the emission line profiles observed in
molecular clouds to an excess of large deviations from the average
velocity difference in the cloud.

\ho\ masers are more direct probes of the velocity field than thermal
molecular lines observed in the cold clouds (see Section~6). Typically,
the available VLBI results provide coordinates and line-of-sight
velocities for $n \approx 100$ maser spots per observation. Thus, there
are $m = n(n-1)/2 \approx 5000$ unique pairings for measuring the
velocity-difference distribution.  Given this relatively large number,
we hoped that the statistics were sufficient to identify possible
deviations from a Gaussian distribution at various spatial scales.
Figures~9 and 10 show examples of velocity-difference probability
distributions obtained for particular spatial ranges in SgrB2(M). The
range for a given spatial scale was chosen to be equal to the scale.
Unfortunately, some individual distributions were found to be not
well-defined, centrally peaked distributions, which is evidently due to
insufficient statistics.

In order to produce a more statistically significant result, we
attempted to co-add the individual distributions. The summation was
done as follows. The velocity difference ($x$ axis) and the number of
pairs or counts ($y$ axis) of the individual distributions were
independently normalized.  The $y$ axis was normalized by dividing all
bin counts by the maximum bin count.  In order to normalize the $x$
axis, we first considered how the absolute value of the maximum velocity
difference of individual distributions scaled as a function of radial
separation between maser pairs.  We found that this function was well
approximated by a power-law. The best fitting straight line gave the
exponent $\approx 1/3$. Given the previously established Kolmogorov
scaling law for the mean absolute velocity differences between maser
points, this result could be anticipated if we assume that the
dispersion of velocity differences at a given scale is proportional to
the mean absolute velocity difference at that scale.  Using the
resulting linear fit, the dispersions of velocity differences for
individual distributions were normalized by multiplying the $x$-values
of a distribution by the ratio
%
\begin{equation}
{{\Delta v_{max}^0}\over {\Delta v_{max}^l}}\;, 
\end{equation}
where $\Delta v_{max}^0$ is arbitrarily chosen to equal the maximum
observed absolute value of the velocity difference occurring in the
central spatial bin, and $\Delta v_{max}^l$ is the maximum absolute
velocity difference value of an individual distribution. Once the $x$
and $y$ axes of the individual distributions were normalized, these
distributions were co-added, producing a single histogram of the
distribution averaged over spatial scales.

The conjoined histogram for SgrB2(M) is shown in Figure~10. It has a
well-defined, centrally-symmetric shape.  However, fitting it with a
single Gaussian results in strong positive residuals in the wings of
the distribution (Figure~10$a$). A two-Gaussian fit, both Gaussians
being centered at zero, results in much smaller residuals
(Figure~10$b$). Due to the method by which this histogram has been
obtained, it contains only averaged (over all the spatial scales)
information about the probabilities of deviations. Comparing the
conjoined histogram with the individual histograms shows that summation
significantly improves the statistics. This is indirect evidence that
the velocity field at all or at most of the accessible spatial scales,
has qualitatively similar statistical properties, including an excess
of large deviations from Gaussian distribution. The two-Gaussian fit of
the conjoined distribution provides a quantitative measure of the
excesses, averaged over the entire scale range. The narrower Gaussian
approximately describes the central part of the distribution, and the
broader one describes its wings.  The ratio ``narrow/broad'' of areas
under these two Gaussians measures the excess of large deviations.
>From Figure~9$b$, this ratio is $\approx 0.63$, considerably less than
unity. Thus, the super-Gaussian wings are significant indeed.

Figure~11 shows the results of the same analysis for four other
\ho\ sources. In all of them, the wings of the distribution are much
broader than those of the Gaussian that fits the central part of the
distribution. This demonstrates that excess of large
velocity-difference deviations is a common feature of the turbulent
velocity fields probed by \ho\ masers.
%
%
%
\section {Implications for Supersonic Turbulence}
In this section we discuss possible implications of our results for the
theory of supersonic turbulence.  The statistical study of \ho\ masers
described in the previous sections reveals three important results: (1)
self-similarity (fractality) of the spatial distribution of the maser
spots, (2) power-law character of the structure functions for the
velocity field traced by the masers, with the power indices close to
their classic Kolmogorov values for incompressible turbulence, and (3)
excess of large fluctuations of the two-point velocity increments. As
all these features are typical of the relatively well studied,
incompressible turbulence, one can suspect that \ho\ masers in regions
of star formation arise in a turbulent medium. If this assumption is
correct, \ho\ masers, due to their record brightness and small angular
sizes, may become unique probes of astrophysical turbulence. Since the
velocity increments at the largest scales in the \ho\ clusters are much
greater than the probable speed of sound in neutral molecular gas, we
deal here, by definition, with the poorly studied supersonic regime of
turbulence. Comparison with the available theoretical knowledge about
incompressible turbulence should therefore be done with high caution.

%
\subsection {Fractal Dimension and Intermittency of Turbulence}

The major conclusion of our analysis in Section~3 is that the spatial
distribution of \ho\ masers is fractal and that the measured fractal
dimension is low, $d\simless 1$. If \ho\ masers trace the dissipation
of supersonic turbulence, we should conclude that the  fractal
dimension of supersonic turbulence is considerably lower than that of
incompressible turbulence, the latter being $\approx 2.6$ (Mandelbrot
1982). If one accepts the hypothesis that mass fluctuations in star
forming clouds are produced by supersonic turbulence (Larson 1995),
then the low fractal dimension ($\approx 1.4$) of a cluster of young
stars in Taurus measured by Larson can be considered as a corroboration
of our conclusion. Another relevant fact may be the observed low
fractal dimension of the large-scale distribution of galaxies
($d\approx 1.2$; Mandelbrot 1982), but the role of turbulence in
shaping this structure is even less clear.

It is helpful to introduce the ``running'' filling factor, $\beta$, of
``daughter'' turbulence elements within ``mother'' elements (Frisch
1995). The dimension of a fractal is expressed through $\beta$ by
%
%
\begin{equation}
d = d_0 - \frac{{\rm ln}\,\beta}{{\rm ln}\,s} \,, 
\end{equation}
where $d_0$ is the dimension of the supporting space and $s<1$ is the
scaling factor from the mother eddies to the daughter eddies.
Constancy of $\beta$ from scale to scale guarantees a well-determined
value of $d$ (a linear plot in Figurs~2 -- 4). From equation~(9)
%
%
\begin{equation}
\beta = s^{d_0 - d}\,, 
\end{equation}
which shows that the filling factor of active daughter eddies in
mother eddies decreases when the fractal dimension of turbulence
decreases. The filling factor of active eddies is a direct measure of
the degree of intermittency (spottiness) of turbulence. We conclude
that highly supersonic turbulence, as revealed by \ho\ masers and
perhaps by the large-scale galaxy distribution and the distribution of
stars in young clusters, is more intermittent than incompressible
turbulence. 

The very possibility of representing the observed spatial distribution
of active turbulence elements by a single power law means that
turbulence is intermittent on virtually all scales. The all-scale
intermittency is also corroborated by the fact that the strong excess
of the large velocity-difference deviations is revealed by the
statistics averaged over different spatial scales (Section~4.2).  This
seems to be an important conclusion, because, in the case of
incompressible turbulence, with its relatively high fractal dimension,
the existence of intermittency in the inertial subrange of the scales
(as opposed to the dissipation subrange) has long been an open question
(Frisch 1995).
%
\subsection {Does Supersonic Turbulence Have an Inner Scale?}
According to the classical work by Kolmogorov (1941b,c, 1942),
incompressible turbulence is characterized by two limiting scales ---
the outer scale $L$, where energy is supplied to the turbulent flow,
and the inner scale $\eta$, where it is dissipated via molecular
viscosity. The ``inertial'' subrange of linear scales $l$, where
kinetic energy is neither injected into turbulence nor dissipated, but
only transferred from larger to smaller scales, is between: $L \gg l
\gg \eta$.  In the inertial subrange, turbulence tends to become
homogeneous and isotropic. Kolmogorov demonstrated that the
inner scale of incompressible turbulence is of the order of
%
%
\begin{equation}
\eta_i \sim \biggl ({{\nu^3} \over {\overline \epsilon}} \biggr)^{1/4}\,, 
\end{equation}
where $\nu$ is the kinematic viscosity of the fluid and $\overline
\epsilon$ is the mean rate of energy dissipation per unit mass, given 
by the equation:
%
%
\begin{equation}
\overline \epsilon = \frac{U^3}{L}\,,
\end{equation} 
where $U$ is the characteristic velocity difference at the outer scale
$L$.  Expression (11) for the dissipation scale is readily obtained from
dimensional considerations; it is the only combination with the
dimensions of length that can be constructed from the two parameters
relevant to this mechanism of energy dissipation, $\nu$ and $\overline
\epsilon$.

Our attempt to derive a possible inner scale for highly supersonic
turbulence is based on two assumptions. First, proceeding from the
common belief that the major mechanism of energy dissipation in
supersonic turbulence is via shock waves, we assume that sonic speed,
$c_s$, rather than molecular viscosity, is the relevant parameter of
the problem.  Our second assumption may be more arguable. We assume
that the second relevant parameter of the problem is the same as for
incompressible turbulence --- the mean rate of energy dissipation,
$\overline \epsilon$. Thus, we assume that $\overline \epsilon$ is an
approximate constant of the energy cascade, equal to the rate, per unit
mass, of supply  of kinetic energy at the outer scale.  In other words,
we assume no significant energy dissipation at intermediate scales.
Although some theoretical and observational arguments can be provided
in favor of this hypothesis (see below), we emphasize that, for the
moment, it is only a hypothesis, whose consequences we would like
to compare with observations.

With these two assumptions, we can derive the dissipation scale for
supersonic turbulence, $\eta_s$ using standard dimensional analysis. It
is easy to show that only one combination can be formed by $c_s$ and
$\overline \epsilon$ with the dimensions of length, namely:
%
%
\begin{equation} 
\eta_s \sim {{c_s^3} \over {\overline\epsilon}}\,.  
\end{equation}
Substituting $\overline\epsilon$ from Equation~(12) into equation~(13), we 
can give $\eta_s$ a more useful form:
%
%
\begin{equation} 
\eta_s \sim {{c_s^3 L}\over {U^3}} = {{L }\over {{\rm M}_L^3}}, 
\end{equation}
where M$_L = U/c_s$ is the typical value of the Mach number associated
with the outer scale. Equation (14) shows that in highly supersonic
turbulence (M$_L\gg 1$), the scale where shock waves begin to dissipate
turbulent energy effectively is many orders of magnitude smaller than
the outer scale $L$.

An argument in favor of our assumption that highly supersonic
turbulence does not dissipate much of its kinetic energy at larger
scales is the observed Kolmogorov, 1/3, slope of the two-point velocity
correlation function in \ho\ masers.  The 1/3 slope is a
straightforward consequence of the conservation of energy during its
cascade along the hierarchy of scales.  Any energy dissipation in the
inertial subrange would produce a steeper slope. The specific kinetic
energy associated with turbulent pulsations on a linear scale $l$ is
$\sim v_l^2$, where $v_l$ is the r.m.s. turbulent velocity on the scale
$l$. This energy is passed to smaller scales in about one ``turn-over''
time,
%
%
\begin{equation}
\tau\sim \frac{l}{v_l}.
\end{equation}
Therefore, the rate of energy transfer is
%
%
\begin{equation}
\overline \epsilon \sim{{v_l^2}\over {\tau}}\sim{{v_l^3}\over {l}}\;.
\end{equation}
If $\overline \epsilon$ is constant, equation (16) gives $v_l\propto 
l^{\frac{1}{3}}$.

Suppose now that some amount of energy is dissipated on each scale.
Then $\overline \epsilon$ is not a constant --- it decreases with
decreasing $l$. Approximating this decrease by a power law,
%
%
\begin{equation}
{{v_l^3}\over {l}}\propto l^{\alpha}\,,\;\; \alpha > 0\,, 
\end{equation}
we have
%
%
\begin{equation}
v_l \propto l^{{{1}\over {3}}(1 + \alpha)}\;, 
\end{equation}
which demonstrates the steeping of the scaling law (since $\alpha > 0$).

Why can energy dissipation in larger-scale shocks be hindered in highly
supersonic turbulence? Here is one possible answer. Although the
potential component of the velocity field (describing compression and
expansion of the gas) should play some part in a supersonic flow, this
part may crucially depend on the unknown boundary and initial
conditions of the flow (e.g. Falgarone \et\ 1994). An extreme case is a
purely vortical initial motion of gas at the larger scales. In this
case, although the velocity increments for these scales (the difference
in velocities of two opposite peripheral points of an ``eddy'') may be
highly supersonic, most converging flows within the eddy, which arise
from fluctuations, will produce {\it oblique} shocks, whose normal
velocity component, $v_{sh}$, will not exceed $c_s$ by much, and thus
it will be small in comparison with the average vortical velocity
increment:  $v_l >> v_{sh} \sim c_s$. Thus, even if a shock of large
scale is formed, the time it needs to sweep the eddy and dissipate its
kinetic energy is much longer than the turn-over time (given in
Equation~15), during which the eddy will disintegrate, passing its
energy to smaller scales.  Only when $v_l$ drops down to $\sim c_s$,
which happens at the inner scale, $v_s$ and $v_l$ become comparable,
and massive dissipation of kinetic energy in random shocks becomes
possible. This gives a possible physical justification to equation (14).

%
\subsection {Intermittency and Kolmogorov Spectrum}
The important conclusion from the results presented in Section~4.1 is
that the exponents of the low-order structure functions for the highly
supersonic turbulence are close to their classic Kolmogorov values. In
particular, the exponents of the second order structure function in all
the investigated sources are close to 2/3. This value was predicted by
Kolmogorov for homogeneous incompressible turbulence at very high
Reynolds numbers (Kolmogorov 1941 a-c). Later, many authors, beginning with
Kolmogorov himself (Kolmogorov 1962), attempted to introduce
theoretical corrections to this exponent which would account for the
experimentally detected intermittency of turbulence (e.g.  Mandelbrot
1967; Frisch, Sulem \& Nelkin 1978; see also Frisch 1995). For example,
in the popular ``$\beta$-model'' of Frisch, Sulem, and Nelkin,
intermittency is assumed to have a fractal geometry, and the corrected
exponent is given by
%
\begin{equation} 
\zeta_2 = \frac{2}{3} + \frac{3 - d}{3}\,,
\end{equation}
where $d$ is the fractal dimension of the set on which intermittent
turbulence concentrates. In the case of incompressible turbulence $d
\approx 2.6 - 2.8$ and, therefore, the codimension $3-d$ and the whole
correction factor [second term in equation (19)] are small.  In fact,
most of these intermittency corrections have historically been
introduced as {\it small} parameters, to account for the small degree of
observed intermittency in incompressible flow. This has made it
difficult to discriminate between different theoretical models, as well
as to judge, in general, the plausibility of the approach treating
intermittency as a ``disturbance'' to the classic Kolmogorov theory.

Highly supersonic turbulent flow should have specific features
different from those of incompressible turbulence. These two regimes of
turbulence should, at least, differ in the ways they ultimately
dissipate energy --- via  shock waves and molecular viscosity,
respectively.  However, the power-law character of the observed
spectrum of supersonic turbulence, over several decades in scale, is a
strong indication that the energy cascade from larger to smaller scales
is as intrinsic a property of supersonic turbulence as it is for
incompressible turbulence. Moreover, the prominent fractal structure of
the set on which supersonic turbulence dissipates its energy and the
non-Gaussian statistics of the two-point velocity increments, as
revealed by \ho\ masers, are strong evidence for qualitative similarity
of both regimes of turbulence also in the sense of an intermittent
character of turbulent activity.

Since the energy cascade and intermittency are the only two assumptions
in the above mentioned models of incompressible turbulence, these
models should likely also work in the case of supersonic turbulence.
The only application of incompressibility of turbulence in these models
is the assumption that energy is not dissipated at the intermediate
(inertial subrange) scales. This assumption is not obvious for
compressible, supersonic turbulence. However, the observed slope of the
two-point velocity correlation function, close to Kolmogorov's,
suggests that dissipation on intermediate scales is insignificant in
the supersonic case too (Section~5.1).

It is important to note, in this connection, that the \ho\ masers
reveal a very low fractal dimension ($d\simless 1$) of the set on which
supersonic turbulence ultimately dissipates in shock waves.  With such
a low $d$, the correction term from the beta-model [equation~(19)] is
large, and we anticipate that, due to intermittency, the slope of the
second order structure function will be at least twice as steep as
Kolmogorov's classic value. Any dissipation of energy on intermediate
scales --- the only possible difference from the incompressible case
--- would steepen the structure function even more. Yet, the observed
structure functions are close to Kolmogorov's.

The observed pronounced intermittency  of turbulence combined with the
classic Kolmogorov velocity structure functions can only be understood
if we accept that intermittency is inherent in turbulence --- not a
mere disturbance of its classic Kolmogorov properties. Such an approach
to incompressible turbulence is being developed by Barenblatt and
Chorin (1997; hereafter B\&C)).  These authors claim, in particular,
that both the tendency of the second order structure function to its
classic dependence on $l$ and an increase of the degree of
intermittency are natural asymptotic properties of turbulence when the
Reynolds number tends to infinity.

As first pointed out by von Weizs\"acker (1951), the Reynolds numbers
of the interstellar gas are, in general, very high. The Reynolds
numbers of the dense nuclei of the star-forming molecular clouds, where
\ho\ masers reside, are especially high, due to the low viscosity of
the dense gas.  The turbulent flows probed by \ho\ masers, have typical
velocities $U \sim 10^6\,$cm/s at a scale of $L \sim 10^{17}\,$cm. With
the typical number density $\sim 10^6\,$\ccm\ and temperature $\sim
10^2\,$K for a molecular cloud core, the kinematic viscosity is $\sim
10^{13}\,{\rm cm^2/s}$, and the typical value of the Reynolds number,
${\rm Re} \equiv LU/\nu$, is $\sim 10^{10}$.

For incompressible turbulence, the Reynolds number is a measure of the
width of the inertial subrange. If the inertial subrange of supersonic
turbulence probed by \ho\ had been limited by viscosity, it would have
been very large.  However, there is little doubt that, in the highly
supersonic regime, the dissipation of energy starts at a much larger
scale than the inner scale determined by viscosity. In Section~5.2, we
postulated the existence in highly supersonic turbulence of an inner
scale determined by energy dissipation in small-scale stochastic shocks
and given by equation (14). Although this inner scale is much larger
than the Kolmogorov dissipation scale ($\eta_s \sim 10^{13}\,$cm;
$\eta_i \sim 10^7\,$cm), there is still a large inertial subrange in
the highly supersonic regime -- almost four decades in projected linear
scale.

One can try to reformulate the B\&C theory and obtain an asymptotic law
for the supersonic regime using the Mach number rather than the
Reynolds number. As is customary in crude phenomenological approaches,
we shall ignore distinctions between structure functions involving
different components of the velocity vectors. Specifically, we shall
assume that the second order structure function $D_{zz}$ involving the
line-of-sight ($z$) component of the difference velocity vector for two
points reflects typical properties of all other structure functions.

Following the reasoning of B\&C, we can assume that in the inertial
subrange of supersonic turbulence
%
\begin{equation}
D_{zz} = f(l, L, \overline\epsilon, c_s)\,,
\end{equation}
where $\overline\epsilon$ is the mean rate of energy dissipation per
unit mass. It is assumed that in this inertial subrange the energy flux
from larger to smaller scales is nearly constant and equal to
$\overline\epsilon$. Applying the standard dimensional analysis, we
find the scaling law for $D_{zz}$:
%
\begin{equation}
D_{zz} = (\overline\epsilon l)^{\frac{2}{3}}\,\Phi\biggl 
(\frac{l}{\eta_s},\, {\rm M}_L\biggr )\,,
\end{equation}
where $\Phi$ is a dimensionless function of its two dimensionless
arguments. We chose the Mach number at the outer scale, M$_L$, and the
running scale $l$ measured in the units of inner scale, $\eta_s$, as
the two dimensionless arguments. In the case of incompressible
turbulence the Reynolds number, instead of the Mach number, enters the
parentheses in equation~(20).

The dimensional analysis, by itself, doesn't tell us anything about the
properties of the function $\Phi$ and its two arguments.  In order to
finalize the scaling law, one must make a {\it similarity assumption}
about the behavior of $\Phi$ as its two arguments tend to infinity.  As
B\&C point out, two different assumptions about this function mark the
historical evolution of incompressible turbulence theory:

1. Complete similarity in both arguments: $\Phi \approx \Phi(\infty, \infty)
=\,$const, when both dimensionless arguments tend to infinity
(Kolmogorov 1941a);

2. Complete similarity in Re, but incomplete similarity in $l/\eta$,
leads to a power-law dependence of $\Phi$ on $l/\eta$ (Kolmogorov
1962).

\noindent
The first assumption leads to the classic Kolmogorov $D(l)\propto
l^{2/3}$. The second assumption introduces an additive correction to
the power index, which has been interpreted as a correction for
intermittency. B\&C argue that the second assumption is internally
contradictory and make the separate assumption of incomplete similarity
in $l/\eta$ and {\it no} similarity in Re. This again generates an
additive correction to the power index, but now the correction depends
on the value of Re. In the limit of Re$\,\to\infty$ the correction
tends to zero and one is left with the classic $D(l)\propto l^{2/3}$
dependence.  The correction is substantial only when Re is not large.

These similarity arguments can be repeated in the case of highly
supersonic turbulence, with the formal substitution of M$_L$ for Re.
The B\&C's theory makes no quantitative estimates of the magnitude of
Re (or M$_L$, in our case) necessary to reach the asymptotic behavior
of $D$.  We can suppose that in the case of turbulence probed by
\ho\ masers, M$_L$ is high enough for the correction to Kolmogorov
value of the power index to lie within the experimental errors.  Thus,
B\&C's theory (and its extension to supersonic turbulence) rejects the
notion that significant intermittency requires a significant correction
to Kolmogorov's 1941 law and thereby reconciles the observed low
fractal dimension (high intermittency) of supersonic turbulence and its
classic Kolmogorov velocity structure functions.
%
%
\section{The Origin of \ho\ Masers}

One can speculate that a jet from a young star produces two basic flow
regimes in the ambient gas: (1) a frontal, high-Mach shock and (2) a
high-vorticity flow due to the velocity shear at the side interface of
the jet and the ambient gas (e.g. Masson \& Chernin 1993). We surmise
that the low-velocity \ho\ masers are associated with the second
regime. The high-velocity \ho\ masers may be connected with the first
regime, but one can anticipate that this connection, and turbulence
produced by the frontal shock, would be more complicated than in the
second regime. Looking for the simplest cases of supersonic turbulence,
we discuss only the second regime and the low-velocity masers in this
investigation.

A power-law spatial-distribution correlation function in all of the
\ho\ sources investigated here signifies self-similar clustering over
almost four orders of magnitude in scale --- from $\sim 10^4\,$A.U. to
$\sim 1\,$A.U. The ``minimal clusters'' in this hierarchy [``features''
in Gwinn's (1994) terminology], are actually spatial and spectral
blends of elementary sources --- those observed through an element of
spectral resolution.  By the order of magnitude, both the elementary
sources and the minimal clusters have a typical size of $\sim 1\,$A.U.,
which is intriguingly close to the predicted dissipation scale of
supersonic turbulence, given by equation (14). In a typical \ho\ maser
source, the low-velocity features are spread over a projected area of
$L \sim 10^{17}\,$cm and occupy a velocity interval $U \approx
20\,$km/s. Taking $c_s\approx 1\,$ km/s (appropriate for probable
kinetic temperatures of several hundred K), we have $M_L \approx 20$
and $\eta_s \sim 10^{17}/(20)^3 \sim 10^{13}\,$cm $\;\sim 1\,$A.U.
Unless this is a coincidence, the smallest clusters of \ho\ masers may
be the sites of the ultimate dissipation of turbulent energy via
stochastic shocks on the inner scale of supersonic turbulence.

This new conceptual approach to \ho\ masers may have several important
consequences for understanding the very mechanism of masing in these
sources. It has often been argued that shock waves provide the best
conditions for pumping \ho\ masers.  Strelnitski and Sunyaev (1973)
interpreted the observed large dispersion of the \ho\ Doppler
velocities in W49N as due to supersonic gas outflow from a young host
star and conjectured that the interaction of the outflow with the
surrounding gas could produce shock waves necessary for pumping.  This
hypothesis has been further developed by many authors, who have
elaborated on details of the shock structure and collisional-radiative
(the first word standing for the source and the second --- for the sink
of the quantum heat engine) or collisional-collisional schemes of maser
pumping behind a shock (e.g., Shmeld, Strelnitski \& Muzylev 1976;
Strelnitski 1984; Kylafis \& Norman 1986, 1987; Hollenbach, McKee \&
Chernoff 1987; Elitzur, Hollenbach \& McKee 1989; Kaufman \& Neufeld
1996). A common feature of all these models is that the pumping shock
is a result of a direct collision of the outflow with a dense blob in
the surrounding quiescent gas, or a direct collision of a dense blob in
the outflow with the surrounding gas (Tarter \& Welch (1986).
In the present model, the pumping energy is not imparted to the masing
gas blobs directly by the stellar wind. Instead, the energy is
channeled to the masers by a cascade from larger scales, which receive
energy from the stellar wind or jets.

In previous models, the shocks pumping \ho\ masers were assumed to have
high velocities. These were either high-speed ($\simgreat 50\,$\kms)
dissociative J-shocks (Elitzur, Hollenbach \& McKee 1989), or slower
($\simgreat 10\,$\kms) C-shocks propagating perpendicular to
the magnetic field (Kaufman \& Neufeld 1996). Two main goals were
pursued in developing those models: (1) the achievement of the maximum
possible abundance of \ho\ molecules via chemical reactions, and (2)
the realization of sufficiently high kinetic temperatures for pumping
the $6_{16} - 5_{23}$ and other masing transitions. It has been argued
(Melnick \et\ 1993; Kaufman \& Neufeld 1996) that to fulfill both
these requirements, temperatures $\simgreat 1000\,$K are needed.  An
important question is whether the slow shocks we advocate in this paper
can provide such temperatures. An analysis of the shock structure is beyond
the scope of the present paper. However, we note that a J-shock
propagating {\it along} the magnetic field lines will have a post-shock
temperature $\simgreat 1000\,$K, if its velocity is $\simgreat 3\,$\kms,
which is a realistic velocity for an inner scale shock in our model.
%
%
\section{Are \ho\ Masers an Adequate Probe of Supersonic Turbulence?} 
If the new conception of \ho\ masers proposed here is correct, they may
become an ideal tool for studying the properties of supersonic
turbulence. In contrast to the large-scale ISM, supersonic turbulence
probed by \ho\ masers has only one source of energy, supplied at the
largest scale --- the interaction of the outflow from a young star with
the surrounding gas. Furthermore, these flows are highly super-virial,
so that gravitational effects are not important.  Most probably, these
flows are also super-Alfvenic.  With the probable magnetic
field strength $B\sim 10^{-3}\,$G in the dense cores of molecular
clouds, and $B\sim 10^{-1}\,$G in the \ho\ maser clumps (Fiebig \&
G\"usten 1989), and with the probable number densities of molecular
hydrogen, $n\sim 10^6\,$ and $n\sim 10^{10}\,$\ccm\, respectively, the
Alfvenic velocity is $v_A = B/(4\pi\rho)^{1/2} \approx B/(4\pi m_{H_2}
n)^{1/2} \approx 2\,$\kms\ in both cases.  This is much less than the
outer scale velocity ($v_L\approx 20\,$\kms). Thus, turbulent
velocities at most of the scales should be greater than the Alfvenic
velocity, which means that the magnetic field does not constrain
turbulent pulsations.  

Masers are more effective than traditional ISM probes of turbulence,
such as thermal or fluorescent spectral lines, from the observational
standpoint. Maser lines are bright and narrow, which allows the spatial
and kinematic structure of the associated flow to be measured with high
precision. Because masing condensations are so small ($\simless 1\,$
marcsec), every spectral feature detected by the interferometer gives a
direct measure of line-of-sight velocity at a given {\it point} in the
flow, projected onto the sky. Indeed, the requirement of the velocity
coherence along the line of sight (to produce lines with the observed
widths $\simless 1\,$\kms), together with the observed line-of-sight
velocity gradients in the transverse direction of $\sim
1\,$\kms\thinspace AU$^{-1}$, and under the assumption that velocity
gradients in the region are more or less isotropic, limits the probable
length of a maser hot spot along the line of sight to $\simless
10\,$AU. This is much less than the size of the whole active region
($\sim 10,000\,$AU) and allows us to consider the \ho\ masers as
point-like  probes of the velocity field, virtually as effective as the
direct probes used to study terrestrial turbulent flows.

Doubts can arise on whether \ho\ masers are an adequate probe of the
geometry of the turbulence dissipation.  For example, due to possible
directivity of their radiation, some masing blobs may be unobserved.
Can this distort the statistics we study? Obviously, the
non-observability of a fraction of places where turbulence dissipates
results in underestimation of the space filling factor of dissipation.
However, this will not affect the deduced fractal dimension of the set
on which dissipation takes place, if all the ``eddies,'' down to the
smallest ones contain {\it some} amount of observable masers. This
follows from equation~(4): reduction of $\sigma$ by any factor (due to
non-observability of a fraction of the masers) will not change the
value of the logarithmic derivative, and thus the value of $d$.

If the directivity of the maser radiation is very high, it can affect
observability of the smallest eddies, containing relatively small
numbers of elementary masers. By an unlucky chance, all the masers
within such an eddy could be turned away from the observer. The larger
eddies, containing more elementary masers, should still be observable,
although their contours will be delineated by reduced numbers of maser
spots.  A lack of observability of the smaller eddies will result in a
decrease of $\beta$ with the decreasing scale [see equation (10)] and
in a corresponding change of the slope of the plots in Figures~2-4.
Using equations (9), (4), and (6), it is easy to convince oneself that
this would steepen the slope of the plot toward the smallest scales in
Figures~2 and 4 and flatten it toward the smallest scales in Figure~3.
These effects are either unseen or quite small in these figures. We
believe, therefore, that the majority of the scales, covering almost
four orders of magnitude, are well represented by the \ho\ masers and
that $d$ is determined adequately, regardless of possible omission of a
fraction of the maser probes caused by radiation directivity.
%
%
\section{Conclusions} 
1. VLBI maps of five \ho\ maser sources in regions of star formation
reveal fractal spatial distribution of the masing hot spots; a
power-law dependence of two-point velocity increments on spatial scale;
and the non-Gaussian statistics of velocity increments (a strong excess
of large deviations from the mean value). All these properties are
known as typical of turbulence.

2. If the \ho\ masers trace turbulence indeed, our quantitative
analysis shows that this highly supersonic turbulence is
characterized by a much lower fractal dimension ($d\simless 1$), and
thus much stronger intermittency, than incompressible turbulence.
Strong intermittency at virtually all the spatial scales is also
confirmed by the excess of large velocity increments at all scales.

3. Unexpectedly, the power indices of the low-order velocity structure
functions for the putative supersonic turbulence are found to be close
to the classic Kolmogorov values for high-Reynolds-number
incompressible turbulence. This is incompatible with the strong
intermittency (low fractal dimension) in traditional approaches to
turbulence, but may find its explanation in the framework of the new
approch put forward by Barenblatt and Chorin.

4. Supersonic turbulence with a high Mach number at its greatest scale
may possess an inner scale, at which the bulk of its energy is
dissipated in low-Mach-number stochastic shocks. The predicted value of
the inner scale is close, by the order of magnitude, to the observed
sizes of the \ho\ hot spots. We hypothesize that the \ho\ masers are
generated in the random shocks at the inner scale of highly supersonic
turbulence produced in the ambient gas by the intensive outflow
from a newly-born star.

{\it Acknowledgements.}  S.G. and B.H. acknowledge with gratitude support
of their participation in this study by the NSF (Research Experiences
for Undergraduates) grant AST 93-0039 and by the Nantucket Maria
Mitchell Association.

\vfill\eject
\centerline{\bf FIGURE CAPTIONS}
Fig.~1. Multi-scale VLBI maps of \ho\ maser source in Sgr B2(M) as
observed on January 23, 1986.  On all maps except the last one, dot
sizes are larger than maser spot sizes.  On the last map, dots are
smaller than observed maser spot sizes ($\sim 1\;$AU, which is perhaps
affected by interstellar scattering). The dots on the last map show
measured positions of spectral channels whose  radial velocities are
indicated near the dots.

Fig.~2. Density-radius fractal measure for \ho\ maser source in
Sgr~B2(M) for the four epochs of observation ($a$ -- $d$) and for their
average ($e$).  The solid straight line  shows linear fit to the data
points (empty circles).

Fig.~3. Box-counting fractal measure for \ho\ maser source in Sgr
B2(M) --- average for the four epochs of observation.

Fig.~4. Density-radius fractal measure for the \ho\ maser sources  
in W49N, W51(MAIN), W51N, and W3(OH).

Fig.~5. Two-point line-of-sight velocity correlation function for
\ho\ masers in SgrB2(M). The empty circles represent data, the straight
solid line shows linear fit. {\it a--d} --- the results for the four
epochs of observation; {\it e} --- the average for the four epochs.

Fig.~6. Two-point line-of-sight velocity correlation function for
\ho\ masers in W49N, W51N, W51MAIN, and W3(OH). 

Fig.~7. Two-point line-of-sight velocity correlation function for
\ho\ masers in one of the two ``clusters'' (streams of bipolar outflow)
in W3(OH).

Fig.~8. Two-point line-of-sight velocity correlation function for a
planar projection of a randomly-filled spherical shell, for different
ratios of expansion and rotation components. Asterisks represent data.
One of the two straight lines in each graph shows a linear fit
(possible only for the largest scales); the other line indicates, for 
reference, a slope of 1/3.

Fig.~9. Histograms of the number of pairs having a given value of the
deviation of their relative velocity from its mean value. $a$:
separation range $18\to 36\,$mas; $b$: separation range $0.32\to
0.63\,$mas; $c$: separation range $0.18\to 0.36\,$mas.

Fig.~10. The probability of deviation of the two-point velocity
difference from its mean value, averaged over scales for the \ho\ maser
spots in SgrB2(M). Both abscissa and ordinate axes are normalized as
described in text. Dots represent data. Solid lines show a
one-Gaussian fit ($a$) and a two-Gaussian fit ($b$) to the data
points.

Fig.~11. Same as in Fig.~10 for \ho\ masers in W49N, W51N, W51MAIN, and
W3(OH) (``Cluster''). Solid line --- observed data, broken line ---
one-Gaussian fit to the central part of the observed distribution.

\newpage
\begin{figure}
\plotone{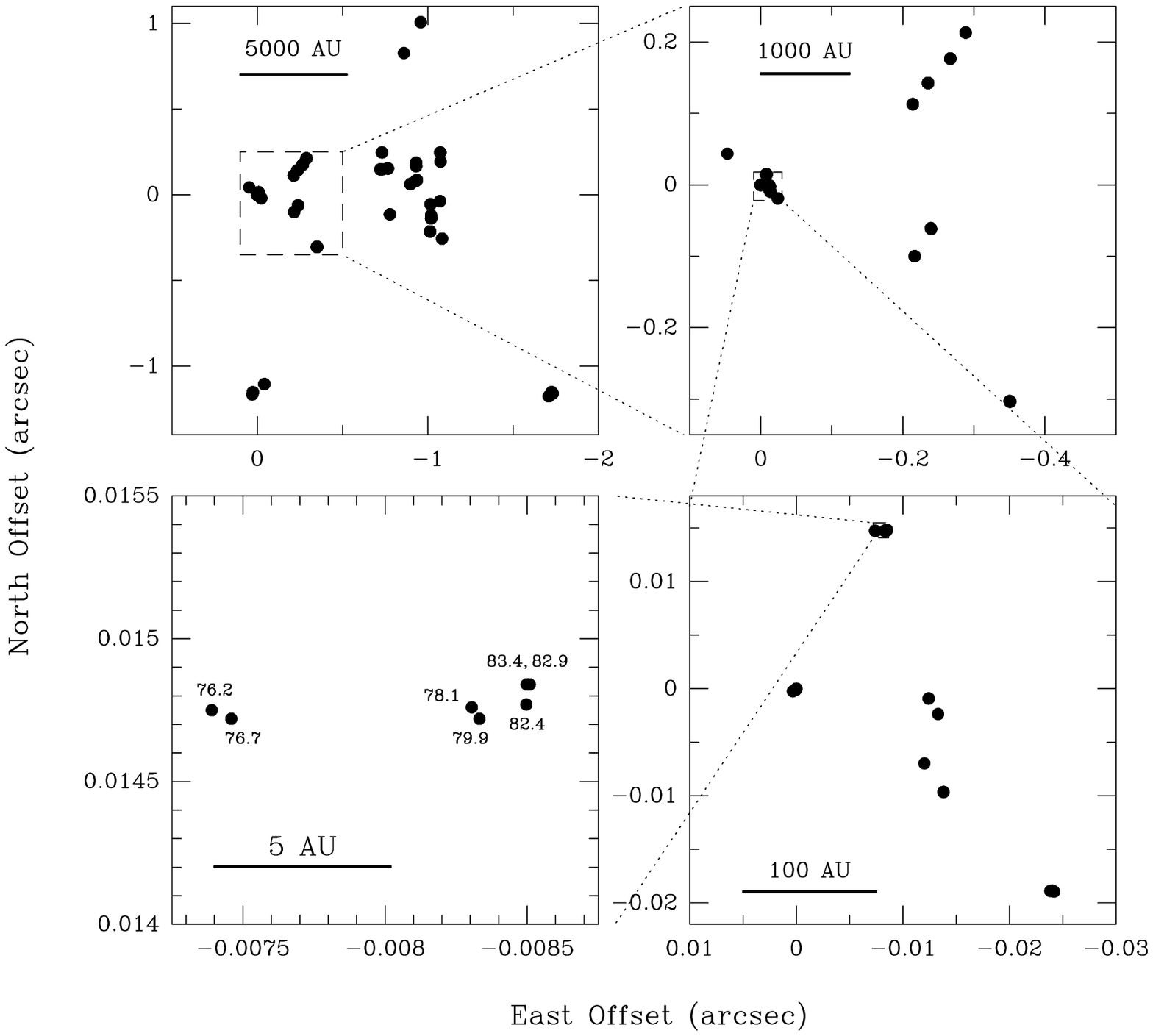}
\caption{}
\end{figure}

\newpage
\begin{figure}
\plotone{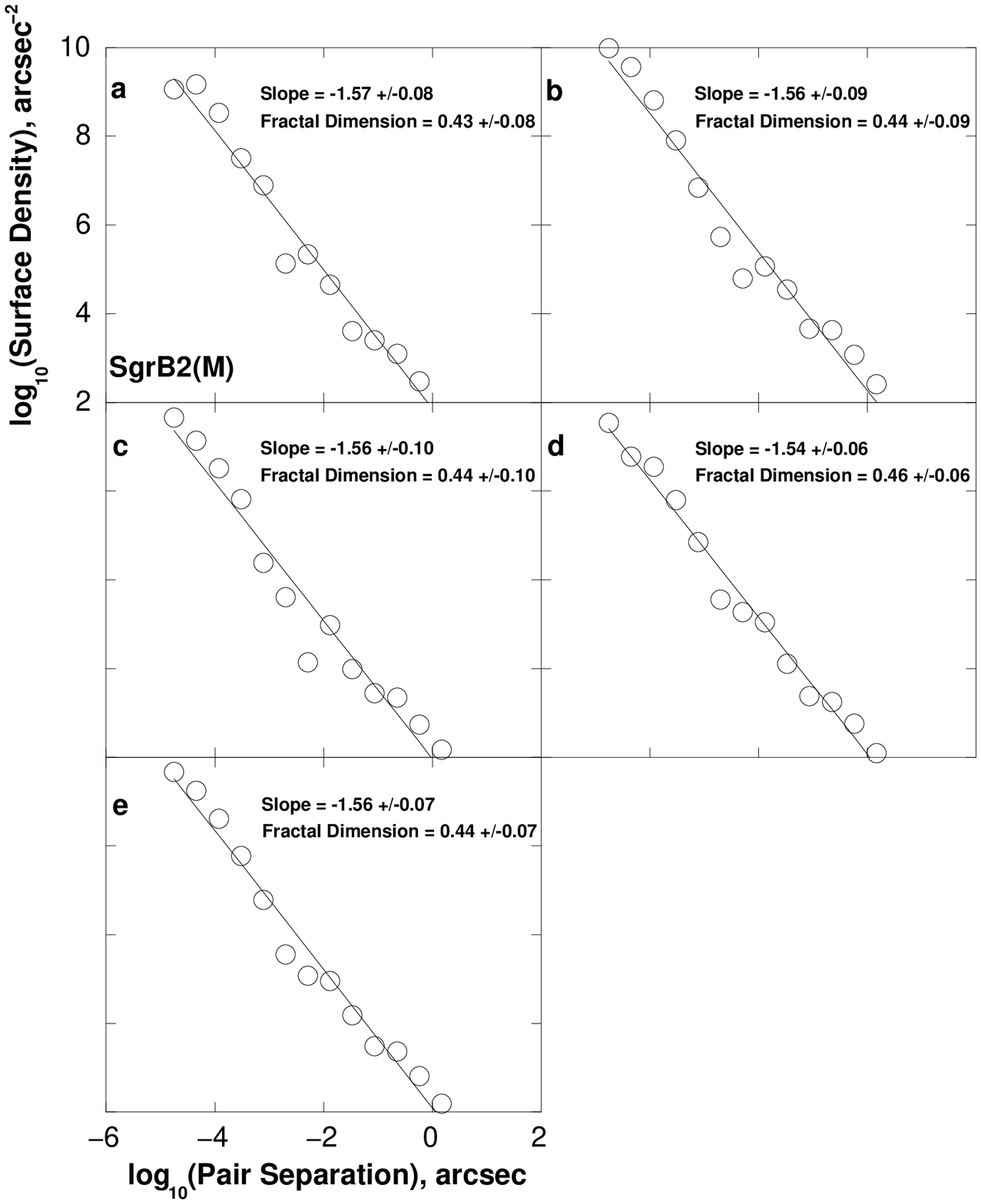}
\caption{}
\end{figure}

\newpage
\begin{figure}
\plotone{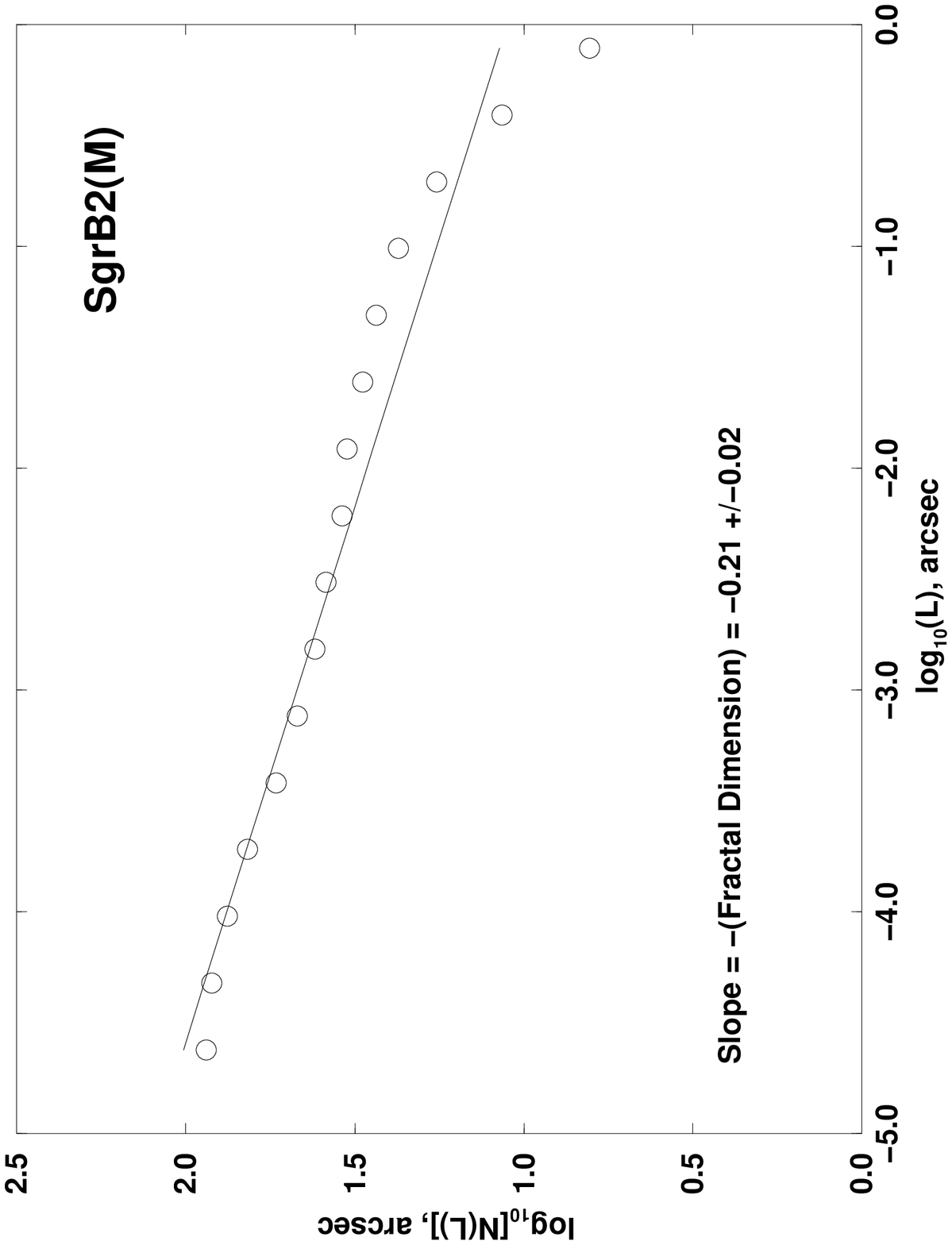}
\caption{}
\end{figure}

\newpage
\begin{figure}
\plotone{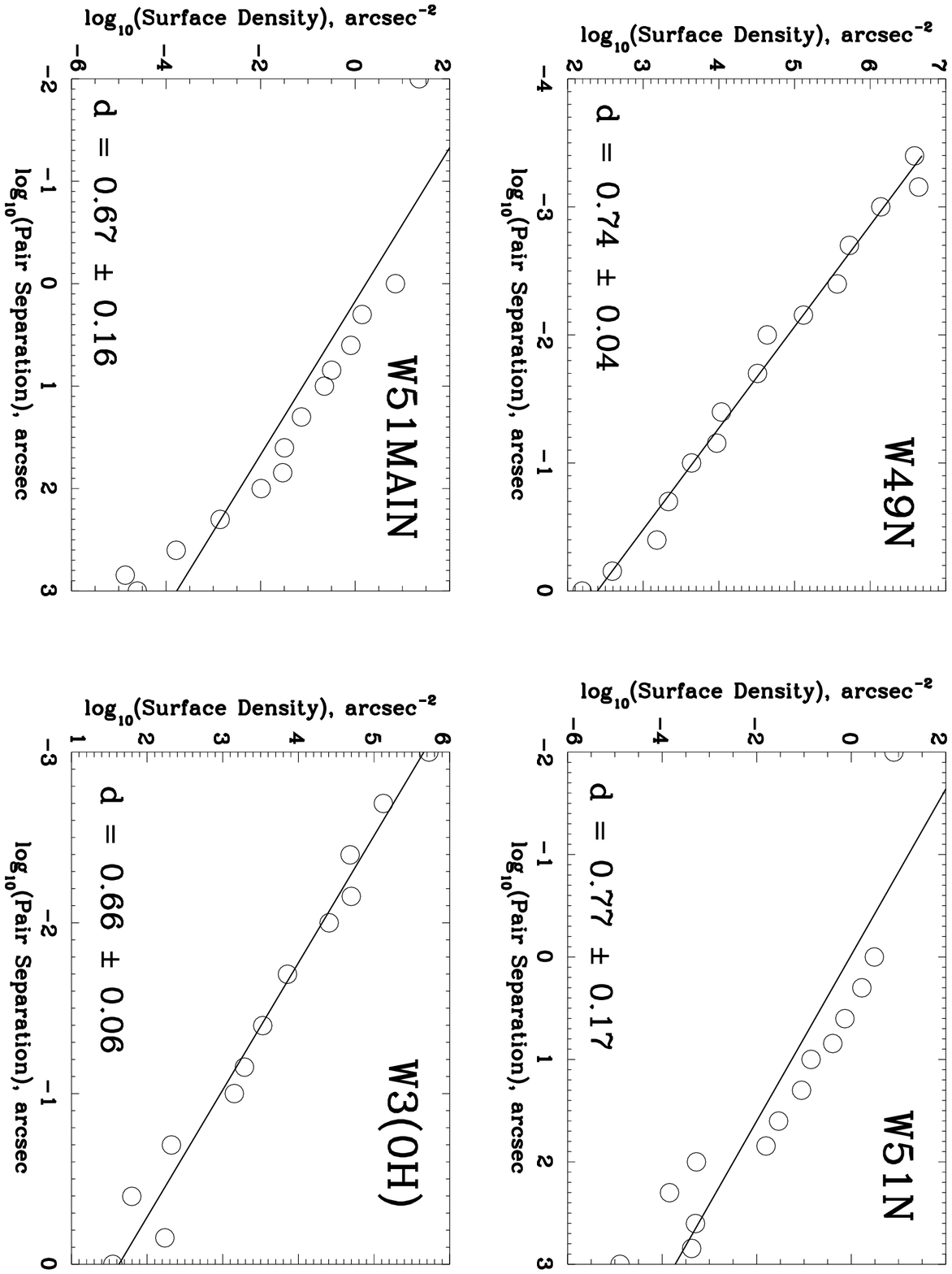}
\caption{}
\end{figure}

\newpage
\begin{figure}
\plotone{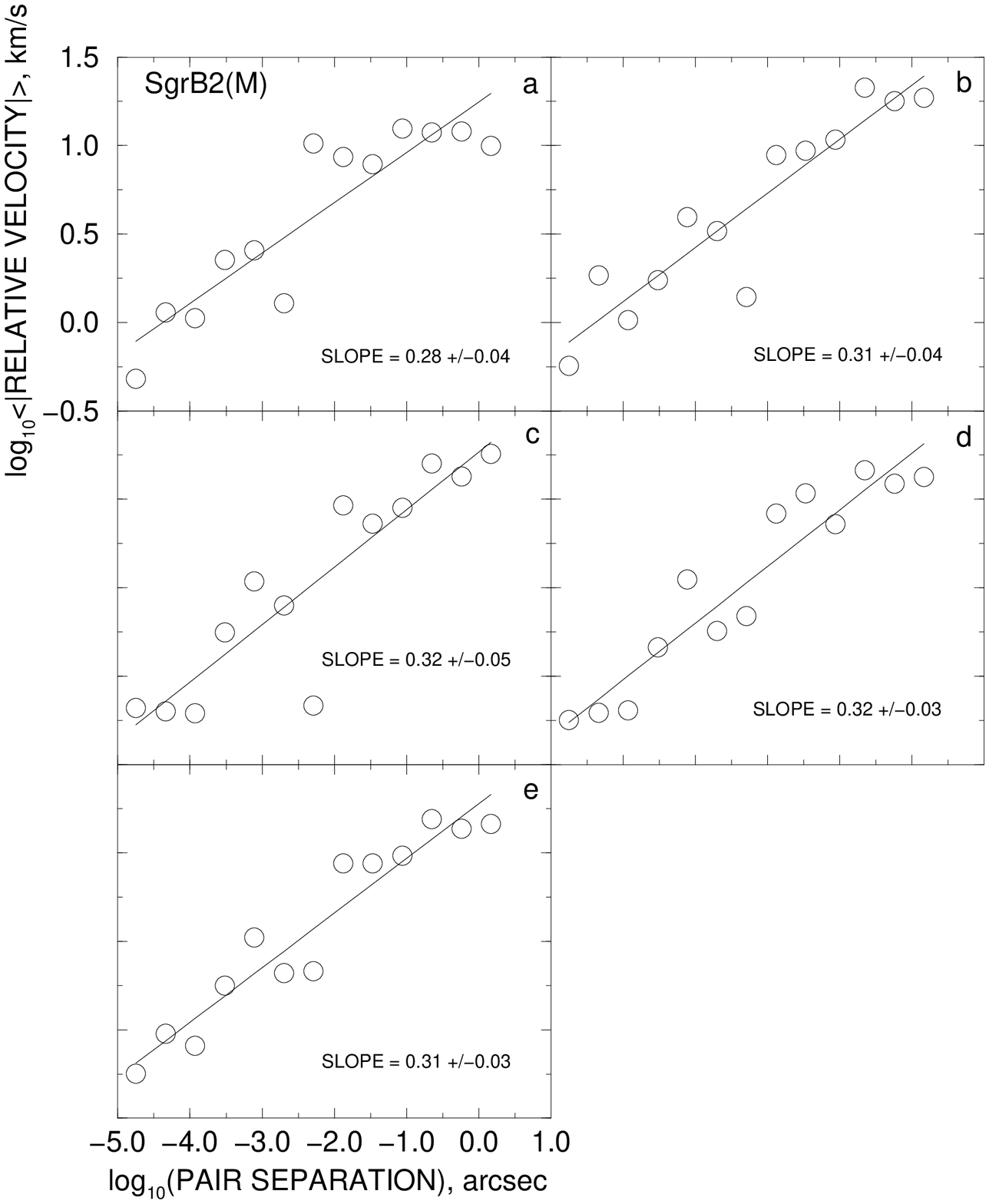}
\caption{}
\end{figure}

\newpage
\begin{figure}
\plotone{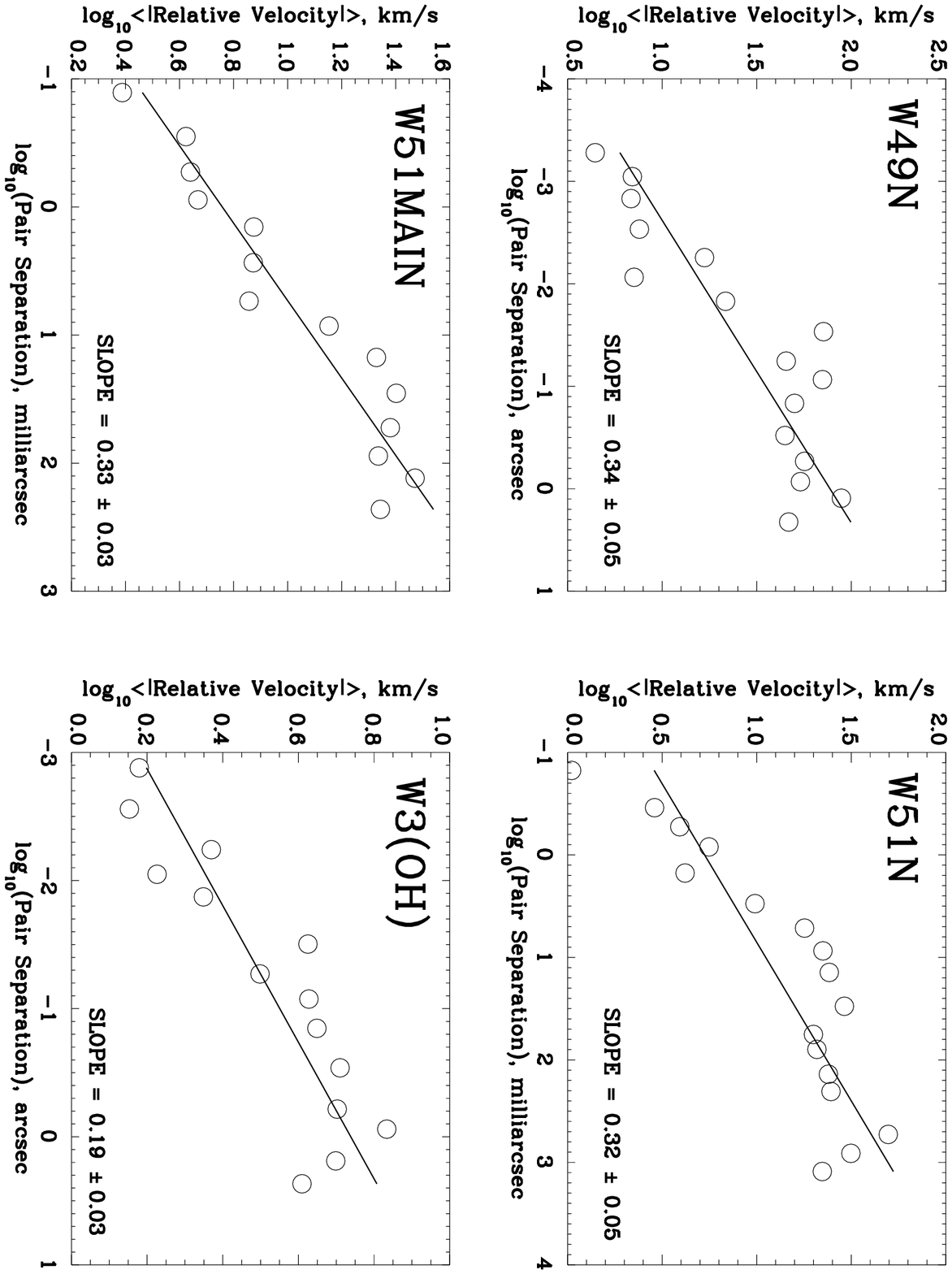}
\caption{}
\end{figure}

\newpage
\begin{figure}
\plotone{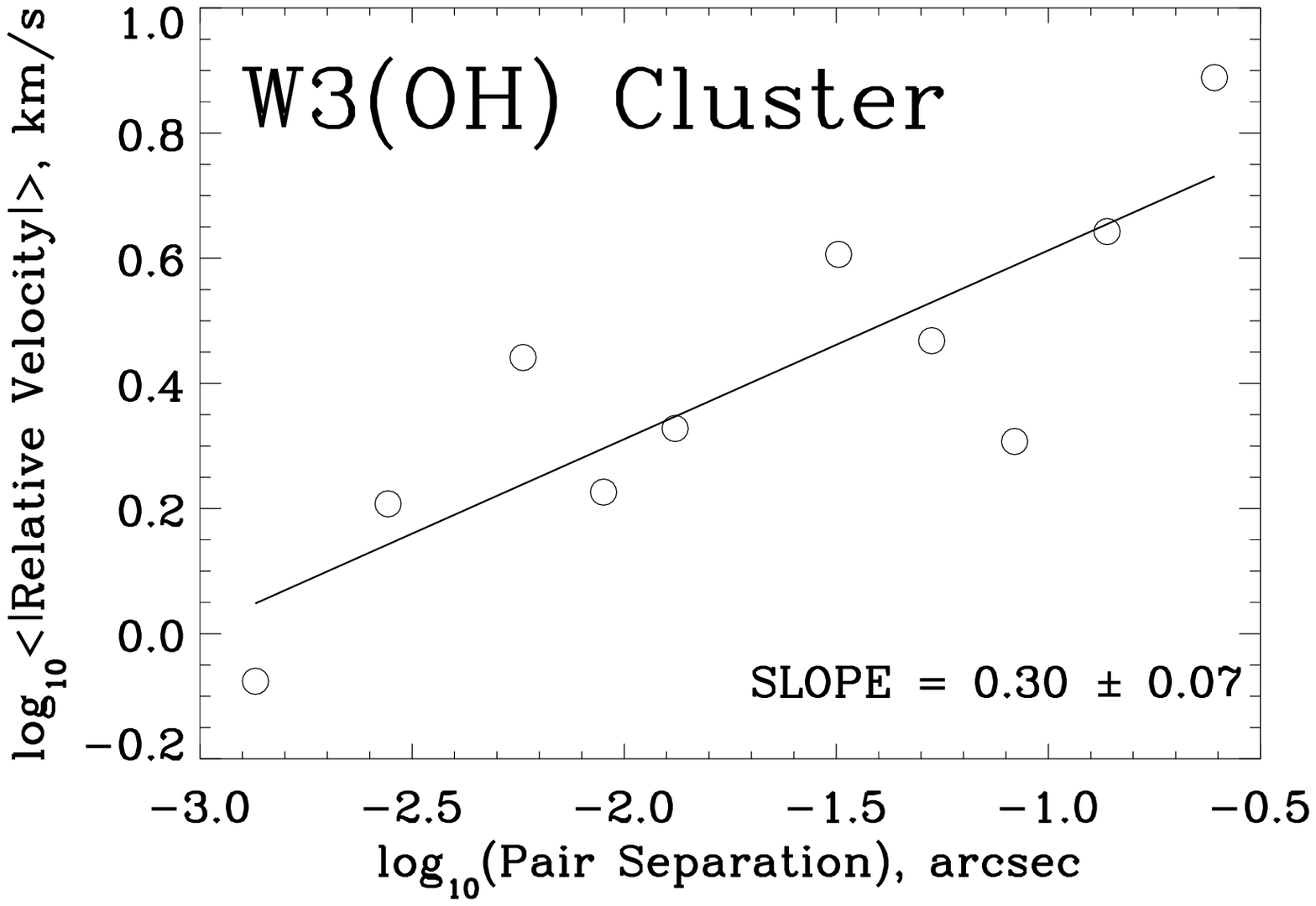}
\caption{}
\end{figure}

\newpage
\begin{figure}
\plotone{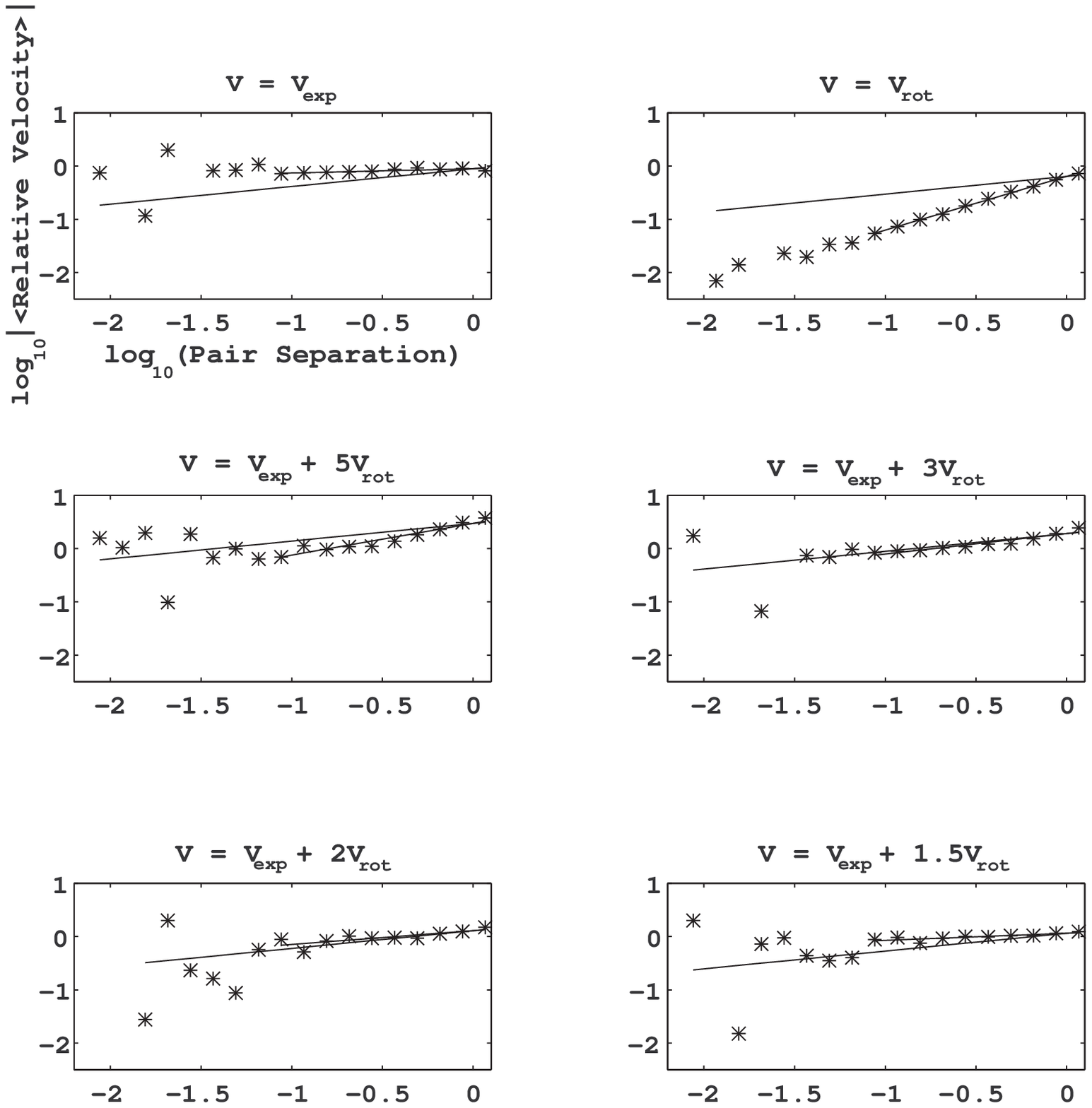}
\caption{}
\end{figure}

\newpage
\begin{figure}
\plotone{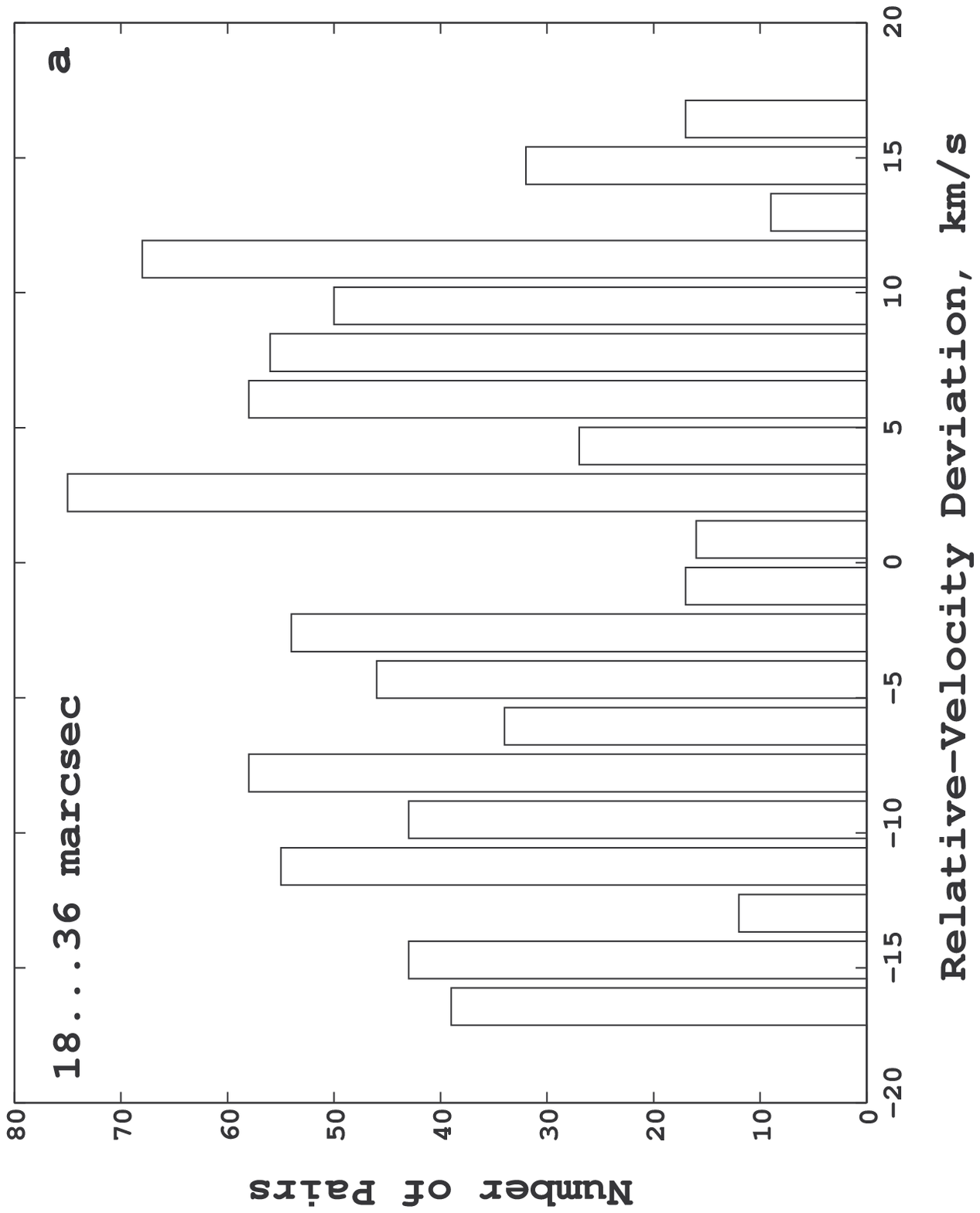}
\caption{}
\end{figure}

\newpage
\begin{figure}
\plotone{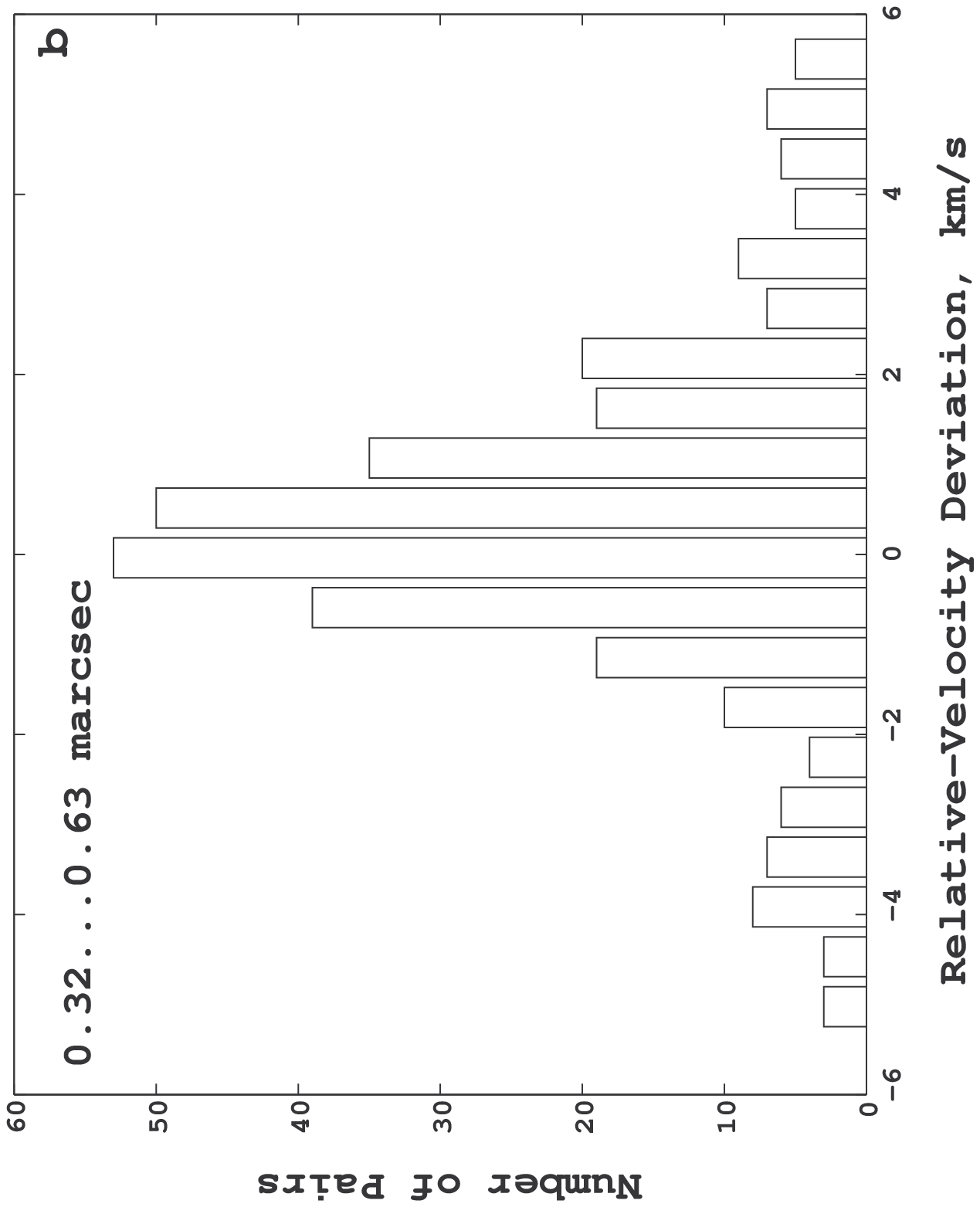}
\end{figure}

\newpage
\begin{figure}
\plotone{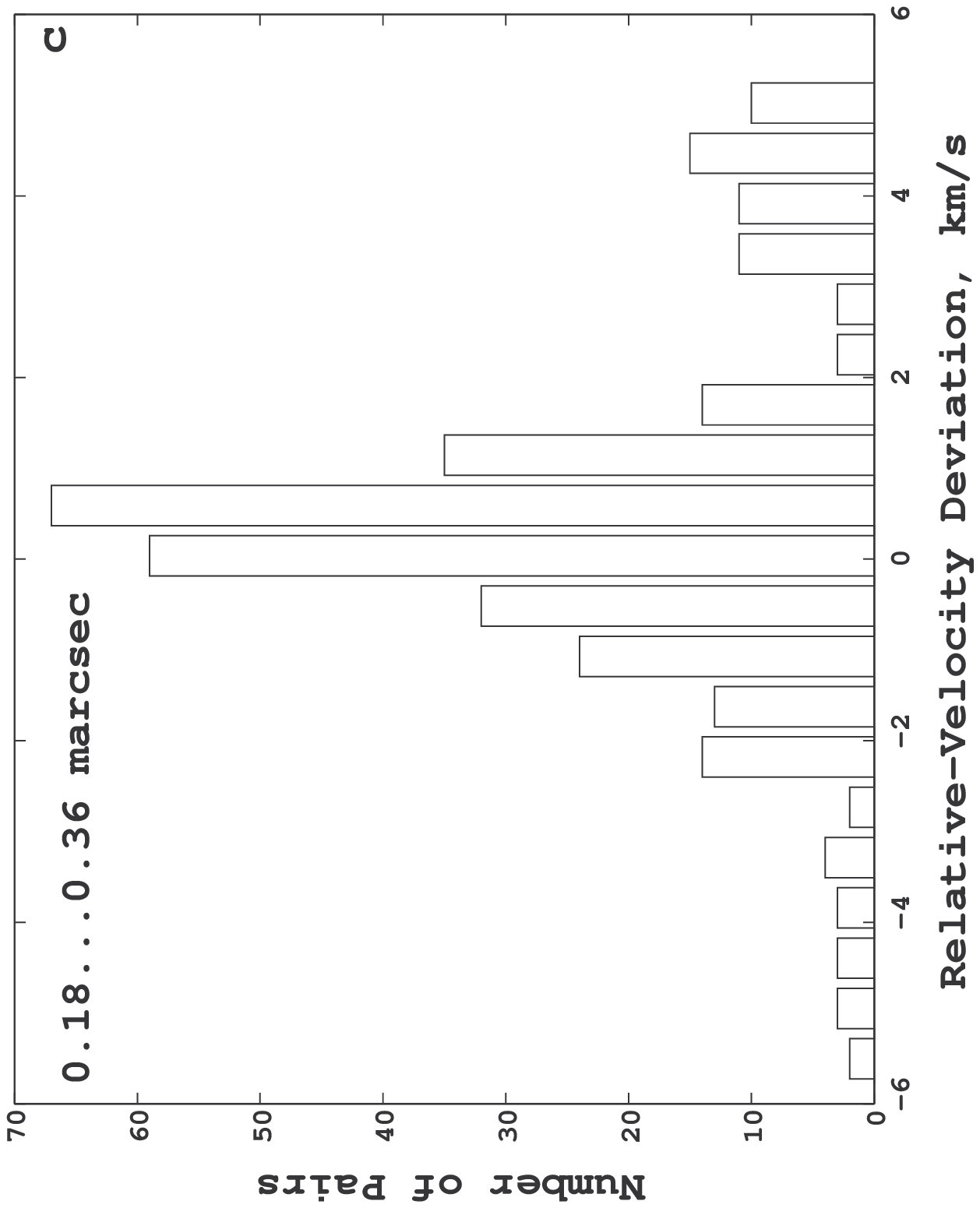}
\end{figure}

\newpage
\begin{figure}
\plotone{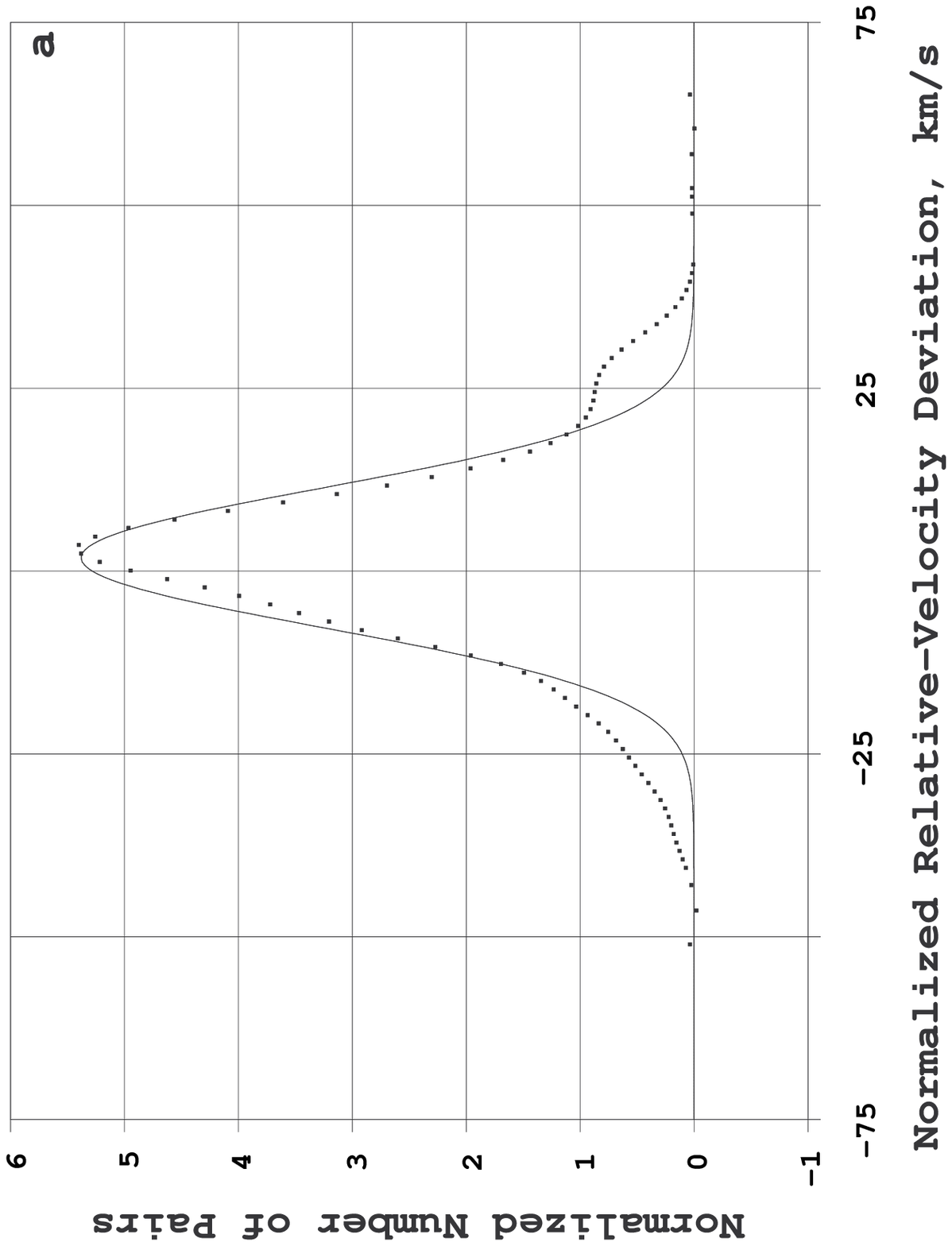}
\caption{}
\end{figure}

\newpage
\begin{figure}
\plotone{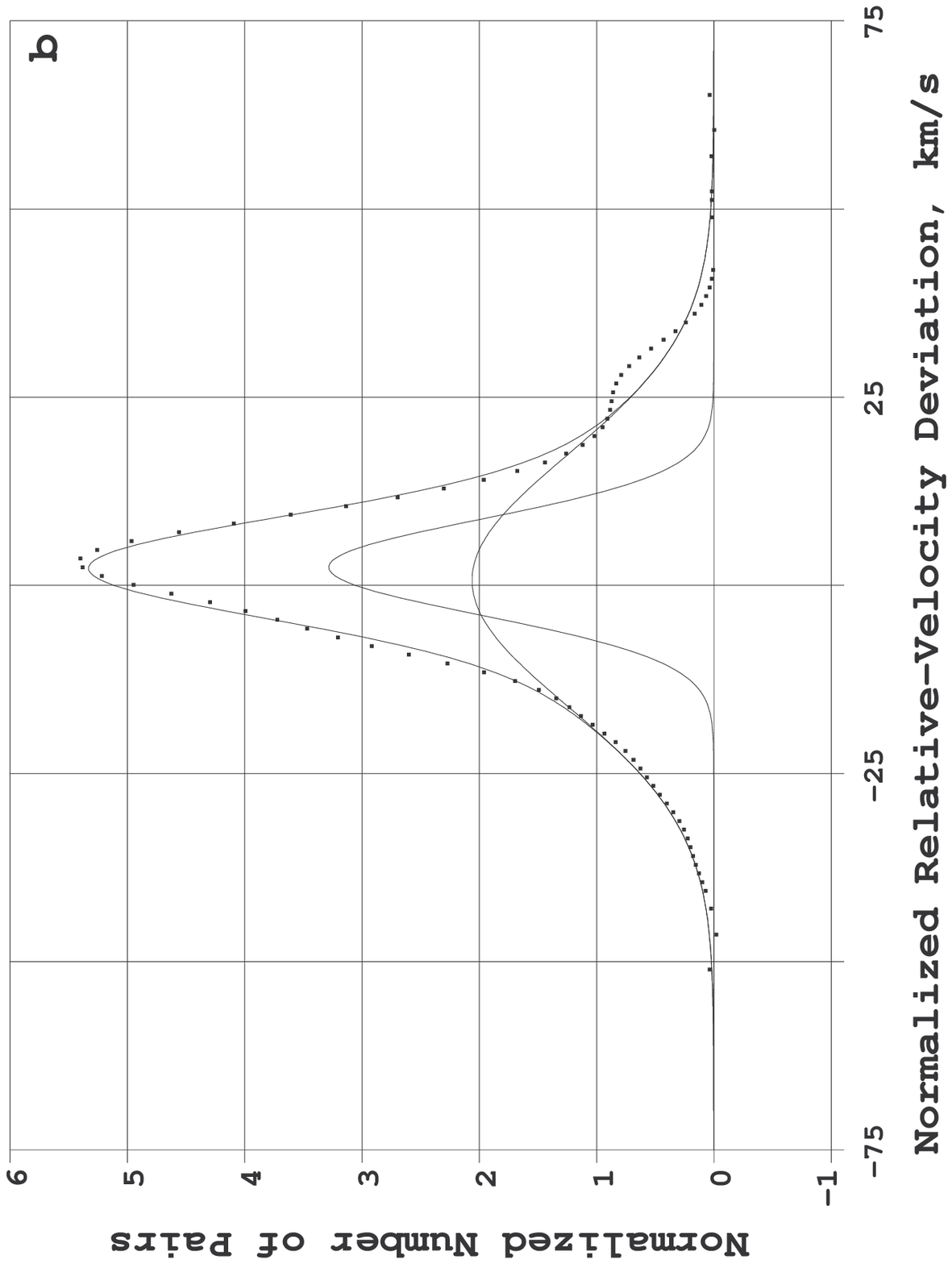}
\end{figure}

\newpage
\begin{figure}
\plotone{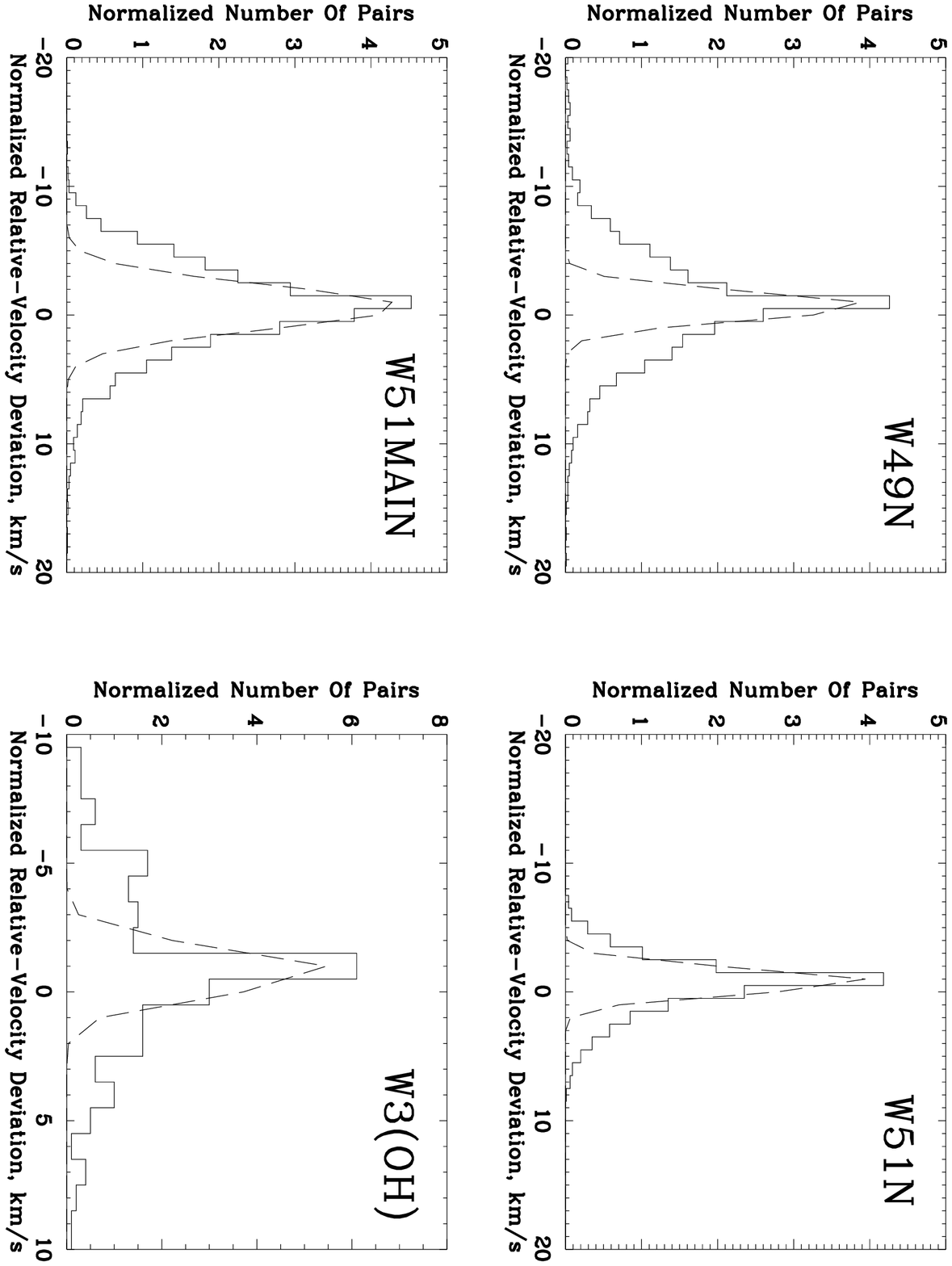}
\caption{}]
\end{figure}
  
\end{document}